\documentclass[fleqn,usenatbib]{mnras}

\usepackage{newtxtext,newtxmath}

\usepackage[T1]{fontenc}

\DeclareRobustCommand{\VAN}[3]{#2}
\let\VANthebibliography\thebibliography
\def\thebibliography{\DeclareRobustCommand{\VAN}[3]{##3}\VANthebibliography}

\usepackage{graphicx}	
\usepackage{amsmath}	
\usepackage{color,soul} 
\usepackage{verbatim}   

\usepackage{tikz,xcolor,hyperref}

\newcommand{\mgii}{Mg\,\textsc{ii}}
\newcommand{\civ}{C\,\textsc{iv}}
\newcommand{\feii}{Fe\,\textsc{ii}}

\newcommand{\lya}{Ly\,\textsc{$\alpha$}}
\newcommand{\hbeta}{H\textsc{$\beta$}}

\newcommand{\mgiil}{Mg\,\textsc{ii}$\lambda$2799}

\newcommand{\hbetal}{H\textsc{$\beta$}$\lambda$4863}

\defcitealias{Onken_2022_QLF}{O22}
\defcitealias{Trakhtenbrot_2011}{T11}

\definecolor{lime}{HTML}{A6CE39}
\DeclareRobustCommand{\orcidicon}{%
    \begin{tikzpicture}
    \draw[lime, fill=lime] (0,0) 
    circle [radius=0.16] 
    node[white] {{\fontfamily{qag}\selectfont \tiny ID}};
    \draw[white, fill=white] (-0.0625,0.095) 
    circle [radius=0.007];
    \end{tikzpicture}
    \hspace{-2mm}
}

\newcommand{\orcidChrisO}{\href{https://orcid.org/0000-0003-0017-349X}{\orcidicon}}
\newcommand{\orcidChrisW}{\href{https://orcid.org/0000-0002-4569-016X}{\orcidicon}}
\newcommand{\orcidSamuel}{\href{https://orcid.org/0000-0001-9372-4611}{\orcidicon}}
\newcommand{\orcidFuyan}{\href{https://orcid.org/0000-0002-1620-0897}{\orcidicon}}
\newcommand{\orcidfanxiaohui}{\href{https://orcid.org/0000-0003-3310-0131}{\orcidicon}}

\title[XQz5 Legacy Sample]{XQz5: A New Ultraluminous z$\sim$5 Quasar Legacy Sample}

\author[S. Lai et al.]{Samuel Lai,$^{1}$\orcidSamuel\thanks{E-mail: samuel.lai@anu.edu.au}
Christopher A. Onken,$^{1,2}$\orcidChrisO\,
Christian Wolf,$^{1,2}$\orcidChrisW\,
Fuyan Bian,$^{3}$\orcidFuyan\,
and Xiaohui Fan$^{4}$\orcidfanxiaohui
\\
$^{1}$Research School of Astronomy and Astrophysics, Australian National University, Canberra, ACT 2611, Australia\\
$^{2}$Centre for Gravitational Astrophysics, Research Schools of Physics, and Astronomy and Astrophysics, Australian National University, Canberra, ACT 2611, Australia\\
$^{3}$European Southern Observatory, Alonso de C\'{o}rdova 3107, Casilla 19001, Vitacura, Santiago 19, Chile\\
$^{4}$Steward Observatory, University of Arizona, 933 N Cherry Ave, Tucson, AZ 85721, USA\\
}

\date{Accepted XXX. Received YYY; in original form ZZZ}

\pubyear{2023}

\begin{document}
\label{firstpage}
\pagerange{\pageref{firstpage}--\pageref{lastpage}}
\maketitle

\begin{abstract}
Bright quasar samples at high redshift are useful for investigating active galactic nuclei evolution. In this study, we describe XQz5, a sample of 83 ultraluminous quasars in the redshift range $4.5 < z < 5.3$ with optical and near-infrared spectroscopic observations, with unprecendented completeness at the bright end of the quasar luminosity function. The sample is observed with the Southern Astrophysical Research Telescope, the Very Large Telescope, and the ANU 2.3m Telescope, resulting in a high-quality, moderate-resolution spectral atlas of the brightest known quasars within the redshift range. We use established virial mass relations to derive the black hole masses by measuring the observed Mg\,\textsc{ii}$\lambda$2799\AA\ emission-line and we estimate the bolometric luminosity with bolometric corrections to the UV continuum. Comparisons to literature samples show that XQz5 bridges the redshift gap between other X-shooter quasar samples, XQ-100 and XQR-30, and is a brighter sample than both. Luminosity-matched lower-redshift samples host more massive black holes, which indicate that quasars at high redshift are more active than their counterparts at lower-redshift, in concordance with recent literature.
\end{abstract}




\section{Introduction}

Radiation released by actively accreting supermassive black holes can be observed across vast distances, making active galactic nuclei (AGN) and quasars some of the most luminous known non-transient astrophysical objects. Due to their potential for extreme luminosities, quasars can be observed at very early cosmic periods, into the epoch of reionisation, the last major phase change of the universe, at $z \gtrsim 6$ \citep[e.g.,][]{Wu_2015, Jiang_2016, Reed_2019, Shen_2019_GNIRS_QSO, Yang_2021, Dodorico_2023_XQR30}. The highest redshift quasars are discovered up to $z \sim 7.5$ \citep[e.g.,][]{Banados_2018, Yang_2020_z7.5, Wang_2021_z7.642}, within the first 700 Myr since the Big Bang. Powered by supermassive black holes (SMBHs), these quasars provide stringent constraints on the size of black hole seeds and their mass evolution \citep[e.g.,][]{Inayoshi_2020, Volonteri_2021, Fan_2022}. 

Large samples of high quality luminous quasar spectra at high redshift can have significant value to the scientific community. Previous large programmes using the Very Large Telescope (VLT) instrument X-shooter \citep{Vernet_2011_Xshooter} have produced high quality echelle spectra of 100 quasars at redshift $3.5 < z < 4.5$ \citep[XQ-100,][]{Lopez_2016_XQ100} and 30 quasars at redshift $5.8 < z < 6.6$ \citep[XQR-30,][]{Dodorico_2023_XQR30}. Other large samples at a similar redshift include 40 quasars at $4.6 < z < 4.9$ with Gemini/NIRI or VLT/SINFONI H-band spectroscopy \citep{Trakhtenbrot_2011} and 50 quasars at $5.6 < z < 6.4$ with Gemini/GNIRS observations \citep{Shen_2019_GNIRS_QSO}. 

Such samples have already been used for a wide variety of science cases. Sub-damped or damped \lya\ systems in the XQ-100 sample were used to study the cosmic density of neutral gas \citep{Sanchez_2016, Berg_2019}. By tracing metal absorbers along the quasar line of sight, these systems have also been used to probe metal abundances in the host galaxy \citep{Perrotta_2016, Perrotta_2018, Berg_2016, Berg_2021}. In addition, cosmological questions were addressed through constraints of the matter power spectrum from the \lya\ forest \citep{Irsic_2017, Yeche_2017}. Similarly, the XQR-30 sample, which is strategically selected from the reionisation epoch, has been the subject of numerous recent studies constraining reionisation models \citep{Bosman_2022, Zhu_2021, Zhu_2022}, measurements of the quasar proximity zone \citep{Chen_2022, Satyavolu_2023}, identification of metal absorption lines \citep{Davies_2023}, and analyses of the black hole environment \citep{Bischetti_2022, Lai_2022}. Further studies are made possible by the public release of spectroscopic data with superior quality and wide spectral coverage.

In this paper, we present ``XQz5'', a new sample of 83 ultraluminous quasars at the redshift range $4.5 < z < 5.3$, an important epoch for studying precursors of modern massive galaxies in the post-reionisation universe. The motivation is to build a sample that is highly complete at the bright end, selecting from the brightest known Southern quasars, which reflect the most massive and rapidly accreting SMBHs. The redshift range was inherited from the sensitivity range of the SkyMapper search which formed the parent sample \citep{Wolf_2020, Onken_2022_QLF}. Similar to XQ-100 and XQR-30, we observed many of our targets with the X-shooter instrument on the VLT, but we supplement our sample with TripleSpec4.1 \citep{Wilson_2004_TripleSpec} near-infrared (NIR) spectroscopic data from the Southern Astrophysical Research Telescope (SOAR). Designed as a legacy sample, this paper is accompanied by the public release of all reduced data products. The high SNR, moderate resolution, and wide spectral coverage of the X-shooter targets suggest that there is potential to exploit the data on the XQz5 targets for similar scientific objectives as XQ-100 and other high-redshift samples. 

The NIR spectroscopic data collected for the XQz5 sample is necessary to study rest-frame UV emission-lines of high redshift quasar, which can be used to constrain metallicities in the broad-line region and black hole properties. It has been observed that in the local universe, black hole and galactic bulge masses are strongly correlated \citep[the $\rm{M}_{\rm{BH}}-\rm{M}_{\rm{bulge}}$ relation; e.g.,][]{Marconi_2003, Haring_2004, Greene_2010}, suggesting co-evolution between host galaxies and their central SMBHs. Characterising the black hole properties at high redshifts is thus valuable for studies of how the galactic and black hole properties came to be tightly correlated \citep[e.g.,][]{Croton_2006, McConnell_2013, Terrazas_2020}. Therefore, in this study, we measure the \mgiil\AA\ emission line which is observed in all 83 targets. Compared to emission lines further in the rest-frame UV which often exhibit blueshifted profiles \citep[e.g.][]{Shen_2016}, the \mgii\ line is typically a better tracer of the galaxy redshift at high redshifts. Using established virial mass relations and bolometric luminosity corrections, we provide robust estimates of the intrinsic black hole properties for the XQz5 sample measured in homogeneous manner.

We have organised the content of the paper as follows: in Section \ref{sec:sample}, we describe our z$\sim$5 quasar sample and observations. In Section \ref{sec:spectral_fitting}, we describe the approach we adopt in modelling the quasar continuum and \mgii\ emission-line. We further discuss how properties of the emission-line can be used to characterise the black hole, using the single-epoch virial mass estimate. In Section \ref{sec:results_discussion}, we discuss our results and contextualise our sample. We summarize and conclude in Section \ref{sec:conclusion}. Throughout the paper, we adopt a flat $\Lambda$CDM cosmology with H$_{0} = 70$ km s$^{-1}$ Mpc$^{-1}$ and $\left(\Omega_{\rm m}, \Omega_{\Lambda}\right) = \left(0.3, 0.7\right)$. All referenced wavelengths of emission-lines are measured in vacuum.

\section{Sample Description and Observation} \label{sec:sample}
Our sample is based on a collection of ultraluminous ($M_{\rm{145nm}} < -27.5$ AB mag) Southern quasars with unprecedented completeness from \citet[][hereafter \citetalias{Onken_2022_QLF}]{Onken_2022_QLF}. Using the SkyMapper Southern Survey \citep[SMSS;][]{Wolf_2018_SkymapperDR1, Onken_2019_SkymapperDR2}, \citetalias{Onken_2022_QLF} identified a sample that is 95\% complete down to $z_{\rm{AB}} = 18.5$ mag, with an overall completeness of $\sim$80\% (See Fig. 4 of \citetalias{Onken_2022_QLF}). The completeness estimates are based on the number of known quasars eliminated by the selection criteria and the linearity of the log source number counts. This implies a 50\% higher completeness at bright magnitudes than a previous reference sample from the Sloan Digital Sky Survey (SDSS) \citep{Yang_2016}. These spectroscopically confirmed quasars contain some of the most massive and fastest growing SMBHs in the early Universe. About $8.5\pm2.9\%$ of the sample can be considered radio-loud, confirming a decrease in the radio-loud fraction with increasing redshift \citep{Lah_2023}. We do not comment on the X-ray properties of the sample, but with the imminent publication of the first data release from the eROSITA All-Sky Survey \citep{Predehl_2021_eROSITA}, a systematic investigation will become possible.

Our sample, with spectroscopic redshifts of $4.5 < z < 5.3$, serves as a high-luminosity extension of the \citet[][hereafter \citetalias{Trakhtenbrot_2011}]{Trakhtenbrot_2011} sample, partially bridging the gap between the XQ-100 Large Programme \citep{Lopez_2016_XQ100} at lower redshifts ($3.5 < z < 4.5$) and the XQR-30 Large Programme \citep{Dodorico_2023_XQR30} at higher redshifts ($5.8 < z < 6.6$). Comparisons of the quasar properties between these samples can reveal evolutionary trends in SMBH growth. We select the brightest quasars from our parent sample while aiming for a uniform distribution across redshift and perform follow-up observations with the spectroscopic instruments: SOAR/TripleSpec4.1, VLT/X-shooter, and ANU2.3m/WiFeS. All of the extracted spectra are made publicly available in a GitHub repository\footnote{\href{https://github.com/samlaihei/XQz5}{https://github.com/samlaihei/XQz5}}.

\subsection{SOAR/TripleSpec4.1}
TripleSpec4.1 \citep{Wilson_2004_TripleSpec} is a fixed-assembly cross-dispersed long-slit near-infrared imaging spectrograph with slit dimensions of 1.1'' by 28'' mounted on SOAR, a 4.1m aperture telescope situated in Cerro Pach\'on, Chile. The instrument covers a simultaneous wavelength range from 0.94 to 2.47 $\mu$m, with spectral resolution R$\sim$3500.

We observed 26 targets in the classical observing mode over 9 nights between 2021-08-22 and 2022-05-15, under the programme IDs 2021B-0036 and 2022A-389756. Using the standard ABBA observing strategy, the total integration time for each target was calculated to achieve a signal-to-noise ratio (SNR) over 20 around the \mgii\ emission-line. For targets in the \textit{H}-band magnitude range $15.8 < \textit{H} < 17.0$, the integration times ranged from 2880s to 12480s between the brightest and faintest targets. We carried out the data reduction and extraction of the infrared spectra using an adaptation of the \texttt{Spextool} \texttt{IDL} package (v4.1) \citep{Cushing_2004_spextool}, specifically designed for TripleSpec4.1. The post-extraction flux calibration and telluric correction were performed with the \texttt{xtellcorr} \texttt{IDL} package \citep{Vacca_2003_xtellcorr} using standard stars observed immediately before and after each target, which are well matched in airmass. The resulting mean $\rm{SNR}_{\rm{2600}}\simeq15.8$ per 40 km s$^{-1}$ pixel measured from the median SNR in the 100\AA\ window between 2550--2650\AA\ in the continuum region near the \mgii\ line. 

\subsection{VLT/X-shooter}
VLT's X-shooter instrument \citep{Vernet_2011_Xshooter} is a multi-wavelength, medium resolution spectrograph with three spectroscopic arms: UVB (300--560nm), VIS (550--1020nm), and NIR (1020--2480nm). We proposed observations for 40 target quasars under the programme IDs 108.22H9.001, 109.23D1.001, and 109.23D1.002. Observations were completed for 32 targets in ``service mode'' between 2021-10-20 and 2022-08-12, 13 of which were also observed with TripleSpec4.1. We also include one target, J2335--5901, which was observed under programme ID 0104.A-0410(A). Under service mode, we defined Observation Blocks (OBs) containing the instrument setup and requested weather conditions.

For the instrument setup, the adopted slit widths are 1.6'' in the UVB, 1.5'' in VIS, and 1.2'' in NIR. Each slit width is the widest slit available for science observations to maximise the signal while providing a nominal resolving power of 3200, 5000, and 4300, respectively. The integration time for each target is calculated to achieve $\rm{SNR} > 20$ around the \mgii\ line in the NIR arm after resampling the extracted spectra into 50 km s$^{-1}$ velocity bins. Of the executed OBs, 71\% were within observing condition specifications with an additional 5\% almost within specifications, resulting in repeat observations of two targets (J091656--251146 and J121921--360933). Where the executed OBs did not meet the desired observing specifications, we manually check the data quality for each individual frame and find that despite the lower quality, all of the data could still be used for further analysis. The total VLT time investment for these new targets is 22 hours. We add two additional targets from the 0104.A-0410(A) programme which invested 0.8 hours of on-target exposure time for J215728.21--360215.1 at $z=4.692$ and 1.6 hours for J233505--590103 at $z=4.53$, resulting in high quality spectra (SNR$_{2600}\sim$40/pixel).

\begin{figure}
	\includegraphics[width=1.0\columnwidth]{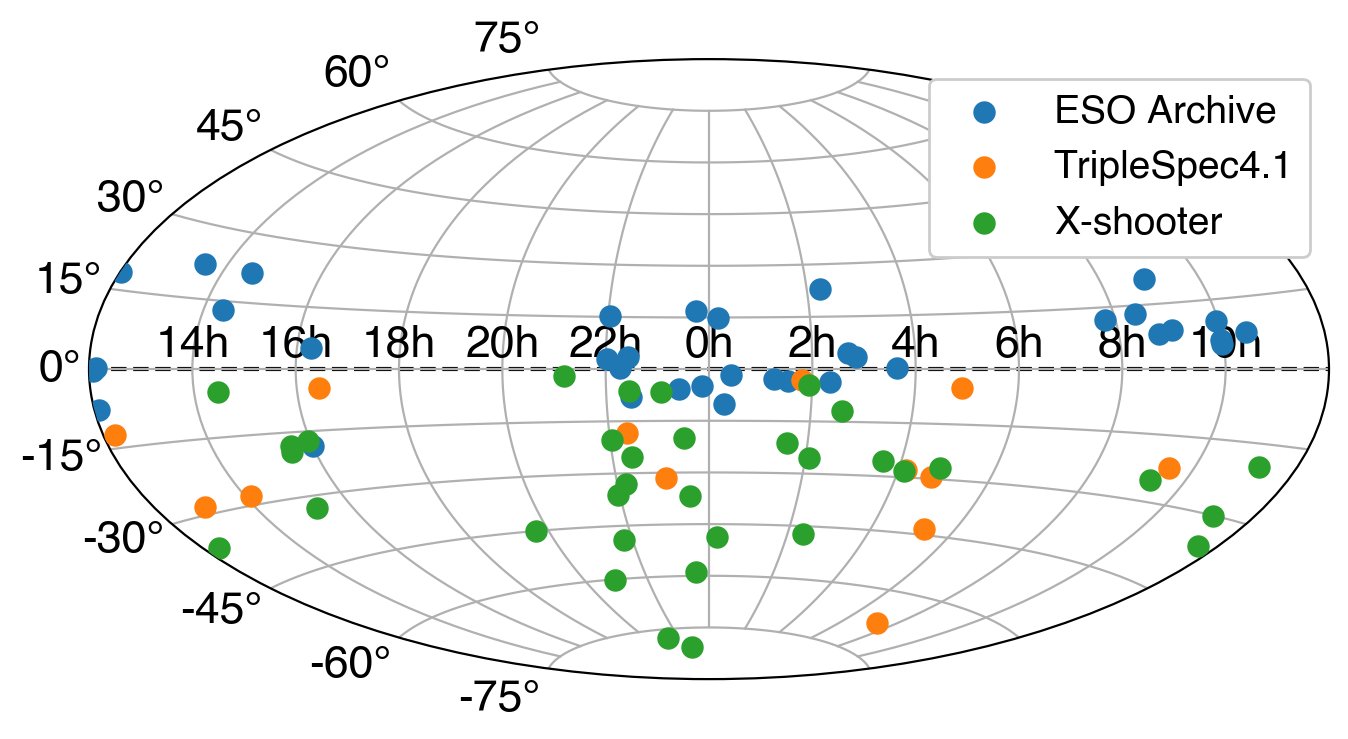}
    \caption{Spatial distribution of the 83 ultraluminous quasars of XQz5 in the equatorial coordinate system. The quasars are colour-coded according to the source of the data presented here. Of the 32 X-shooter targets marked in green, 13 are also observed with TripleSpec4.1 and one (J222612--061807) was later found to have existing data in the ESO Archive.}
    \label{fig:aitoff_skydist}
\end{figure}

We further supplement our sample with additional X-shooter data from the ESO Science Archive Facility of 36 targets, originating from the programmes 084.A-0574(A), 084.A-0780(B), 087.A-0125(A), 094.A-0793(A), 098.A-0111(A). We selected these additional targets from the Million Quasars (Milliquas) v7.7 Catalogue\footnote{\href{https://quasars.org/milliquas.htm}{https://quasars.org/milliquas.htm}} \citep{Flesch_2021_Milliquas}, most of which are bright quasars in the redshift range $5.0 < z < 5.3$ concentrated at more northern declinations as seen in Figure \ref{fig:aitoff_skydist}. The targets from 084.A-0574(A), 087.A-0125(A), and 094.A-0793(A) were observed with slit dimensions such that the effective resolution is R = 8900, 5600 for the VIS and NIR arms, respectively. Targets from 098.A-0111(A) and 0100.A-0243(A) were observed with higher resolution NIR arm, where R = 8900, 8100, and BR 1202--0725 from 084.A-0780(B) was observed with R = 11400, 8100.

Due to the high sky-background level, extraction of the NIR spectra can be non-trivial. Although ESO provides a pipeline for X-shooter data reduction \citep{Freudling_2013_ESOReflex}, we find that the post-extraction spectra often exhibited significant sky-subtraction residuals. Consequently, we opted to use the semi-automated reduction procedure implemented in the \texttt{PypeIt} package \citep{Prochaska_2020_Pypeit}, which contains code infrastructure for X-shooter data reduction. The initial flux calibration is derived from a standard star typically observed within a day of the observation. However, the standard is not often matched to the airmass of our targets, so we later rescale the spectra to the observed photometric magnitudes as described in Section \ref{sec:post-processing}. Furthermore, the telluric correction algorithm in \texttt{PypeIt} utilises a large telluric grid to find the best-fitting atmospheric model in place of deriving the correction from a telluric standard. 

After the extraction of the 1D spectra of all targets observed with X-shooter, the resulting mean $\rm{SNR}_{1450} \simeq \rm{SNR}_{2600}\simeq16.8$ per pixel measured from the median SNR between 1400--1500\AA\ and 2550--2650\AA.

\subsection{ANU2.3m/WiFeS}
The Wide Field Spectrograph (WiFeS) instrument \citep{Dopita_2007_WiFeS, Dopita_2010_WiFeS} installed on the fully automated Australian National University 2.3m telescope is a double-beam integral field spectrograph with a wavelength coverage of 330--930 nm. For this sample, WiFeS observations are conducted on the 13 targets without optical coverage, using R = 3000 gratings.
The data reduction is performed with the standard Python-based pipeline, \texttt{PyWiFeS} \citep{Childress_2014_PyWiFeS} and extraction of the 1D spectra from the IFU data cubes was performed with \texttt{PySpecExtract}\footnote{\href{https://github.com/nishamrutha/PySpecExtract}{https://github.com/nishamrutha/PySpecExtract}}. The resulting mean $\rm{SNR}_{1450} = 6.1$ per 50 km s$^{-1}$ pixel. We re-calibrate the flux normalisation using optical photometry with the same method applied for X-shooter data as described in the folllowing Section \ref{sec:post-processing}.
\\

In Table \ref{tab:XQz5_sample}, we list 
properties of all targets and 
extracted spectra in our sample. The columns from left to right 
are the target name, coordinates, redshift derived from the peak of the \mgii\ emission-line (the details of which is presented in Section \ref{sec:spectral_fitting}), programme ID, Milky Way extinction colour excess (see Section \ref{sec:post-processing}), magnitudes in the \textit{i, z, H}-bands and signal-to-noise ratios in 100\AA\ windows centered at 1450 and 2600\AA. The optical \textit{i} and \textit{z}-band photometry are obtained from SMSS DR4 \citep{Onken_SMSSDR4}, if available, or Pan-STARRS DR1 \citep{PanSTARRS}. The near-infrared \textit{H}-band photometry is from one of various surveys: VISTA Hemisphere Survey \citep[VHS;][]{VHS} DR5/6, VISTA Kilo-degree Infrared Galaxy Survey \citep[VIKING;][]{VIKING} DR2, UKIRT Infrared Deep Sky Survey \citep[UKIDSS;][]{UKIDSS} DR9, or Two Micron All-Sky Survey \citep[2MASS;][]{2MASS}. In some cases, the \textit{H}-band magnitude is interpolated (using the relations from \citetalias{Onken_2022_QLF}) from the VHS \textit{J} and \textit{K}-bands or the SMSS/Pan-STARRS \textit{z} and ALLWISE/CatWISE2020 \textit{W1} bands \citep{WISE, ALLWISE, Marocco_2021_catwise2020}. The survey sources of the optical and near-infrared photometry are indicated in the supplementary table. The full table, extended with associated uncertainties, is available online as supplementary material.

\subsection{Post-Processing} \label{sec:post-processing}
Our combined sample consists of 83 ultraluminous quasars based on new observations from SOAR/TripleSpec4.1 and VLT/X-shooter alongside additional X-shooter data from the ESO Science Archive Facility. These quasars are found in the redshift range $4.5 < z < 5.3$ with \textit{H}-band magnitudes $15.44 < \textit{H} < 18.83$ and their sky distribution in equatorial coordinates is presented in Figure \ref{fig:aitoff_skydist}. Here, we describe the post-processing steps taken to clean the spectra and prepare them for continuum and emission-line modelling.

\begin{enumerate}
    \item We transform the spectrum to the rest-frame by estimating an initial redshift from the peak of the observed \mgii\ data. We then apply a per-pixel SNR floor of 5 and mask all data below. 
    \item We apply a sigma-clipping mask across box-widths of 40 pixels, using a 3$\sigma$ threshold to remove narrow absorption features and noise above 3$\sigma$.  Narrow absorption features, if left in the spectrum, can affect the modelling of the quasar continuum or intrinsic line profiles of broad emission lines. 
    \item The spectra are resampled from their original post-extraction resolution (13.5/40 km s$^{-1}$ for X-shooter NIR/TripleSpec4.1 and <11.6/50 km s$^{-1}$ for X-shooter VIS/WiFeS) into a common wavelength grid from rest-frame 1050--3600\AA\ using a flux-conserving algorithm, \texttt{SPECTRES} \citep{Carnall_2017}, with a velocity resolution of 50 km s$^{-1}$ for each bin. Rebinning and consolidating the flux in 50 km s$^{-1}$ bins increases the SNR of the spectral data, most notably for the X-shooter data, and standardises the output from different spectroscopic instruments for continuum and emission-line modelling.
    \item As the initial X-shooter flux calibration is based on a standard that is not reliably observed in the same night or under similar conditions, we independently recalibrate the X-shooter spectra to photometry. We integrate the observed spectroscopic data over the bandpass transmission curves obtained from the SVO Filter Profile Service\footnote{\href{http://svo2.cab.inta-csic.es/theory/fps/}{http://svo2.cab.inta-csic.es/theory/fps/}} \citep{SVO_Filter_Profile_Service} and scale the flux to match the observed magnitudes. For the X-shooter VIS arm, we use the inverse variance to weight the flux calibration between the \textit{i}-band and \textit{z}-band. We then scale the X-shooter NIR arm according to the calibration of the VIS arm, enforcing continuity between arms. Targets observed with TripleSpec4.1 are already calibrated to an airmass-matched standard observed immediately before or after the quasar, so we take its original calibration to be most reliable. If there is TripleSpec4.1 near-infrared spectroscopic data available for an object, then we use the TripleSpec4.1 spectrum to scale all other spectroscopic data. If we calibrated TripleSpec4.1 spectra to \textit{H}-band photometry taken at a variety of different epochs, we find that the median absolute correction for the TripleSpec4.1 sample to be 15\% of its original flux calibration with an additional 15\% RMS deviation. In some cases, prior to the rescaling and splicing of the optical and near-infrared spectra, a discontinuity can be observed from the combined effects of chromatic slit losses and quasar variability. We note that the 15\% median absolute calibration discrepancy would result in $<0.1$ dex uncertainty in the estimated continuum luminosity.
    \item We correct for Milky Way extinction by adopting $R_{\rm{v}} = 3.1$ with the \citet{F99_Dust} extinction curve and colour excess from the Schlegel, Finkbeiner \& Davis \citep[SFD;][]{Schlegel_1998} extinction map. We then apply a suggested 14\% recalibration to the colour excess, $E(B-V) = 0.86 \times E(B-V)_{\rm{SFD}}$, informed by fits to the blue tip of the stellar locus \citep{Schlafly_2010}. We assume no further dust extinction from the quasar host galaxy, following the common practice for similar studies of quasar line properties, including \citet{Shen_2011} -- which defines our mass calibration (see Section \ref{sec:virial_mbh}) -- and others \citep[e.g.,][]{Rakshit_2020, Wu_2022, Lai23_XQ100}. Furthermore, luminous quasars are intrinsically less likely to exhibit notable dust extinction, with the majority of quasars consistent with null colour excess \citep[e.g.,][]{Krawczyk_2015}. Due to the ultraluminous nature of the quasars in this sample, we do not account for the starlight contribution in the spectra, which is insignificant compared to the AGN. 
\end{enumerate}

\begin{figure}
	\includegraphics[width=1.0\columnwidth]{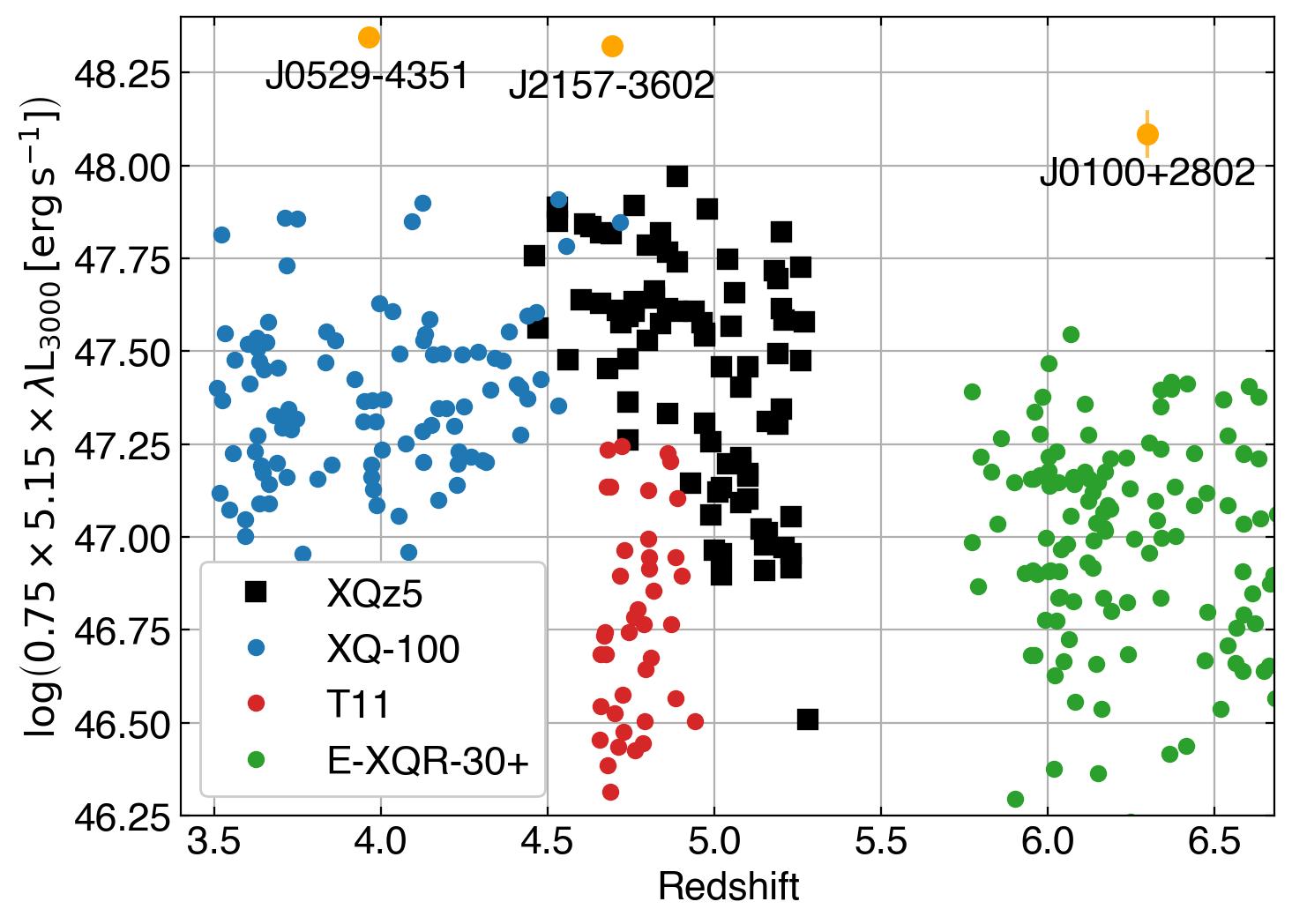}
    \caption{Our sample, XQz5 (black), flanked by two other X-shooter programmes, one at lower redshift, XQ-100 (red), and one at higher redshift, E-XQR-30+ (orange). XQz5 serves as a high luminosity extension of the SDSS-selected \citetalias{Trakhtenbrot_2011} sample (green) at a similar redshift, where the luminosity is measured from the 3000\AA\ monochromatic flux of the power-law continuum. The three named targets are the brightest known quasars in the three redshift ranges, with J0529--4351 and J2157--3602 as the current first and second most luminous known quasars, respectively.}
    \label{fig:sample_comparison}
\end{figure}

After post-processing of our sample, we find the median $\rm{SNR}=27.2$ per 50 km $s^{-1}$ velocity bin measured from the median SNR between 2700--2750\AA\ and 2850--2900\AA. In Figure \ref{fig:sample_comparison}, we show our sample, XQz5, compared to other high-quality X-shooter programmes, XQ-100 and Enlarged XQR-30 Plus (E-XQR-30+), which are concentrated on lower and higher redshifts, respectively. The E-XQR-30+ sample is composed of the core 30 quasars from XQR-30 and supplemented with high-redshift literature quasars with NIR spectra as described in \citet{Dodorico_2023_XQR30}; in addition for this paper, we have extended E-XQR-30+ with 112 quasars listed in the \citet{Fan_2022} review paper. We note that the  gravitationally lensed object QSOJ0439$+$1634 has been removed. We also show quasars from \citetalias{Trakhtenbrot_2011}, which is a lower luminosity sample at a similar redshift range, and we plot three notable quasars: J0529--4351 \citep{Onken_2023_AllBRICQS}, J2157--3602 \citep{Wolf_2018_J2157, onken_2020_J2157}, and J0100$+$2802 \citep{Wu_2015}, the three most luminous quasars in the respective redshift ranges of each sample, of which we include J2157--3602 in our sample. Other luminous quasar samples such as ELQS \citep{Schindler_2017_ELQS} and QUBRICS \citep{Cristiani_2023_QUBRICS} also reach up to $z\sim5$, but they have not been observed with a comparable broad spectral coverage as the X-shooter samples, which include observed-frame near-infrared.

We provide all of the observed-frame telluric-corrected spectra in the supplementary material. The spectra are provided in the native resolution and do not include the post-processing steps described in this section, except for the flux re-calibration. Raw data files are available in their respective data archive facilities, the ESO Science Archive Facility\footnote{\href{http://archive.eso.org/cms.html}{http://archive.eso.org/cms.html}} and NOIRLab Astro Data Archive\footnote{\href{https://astroarchive.noirlab.edu/}{https://astroarchive.noirlab.edu/}}.

\begingroup
\begin{table*}
\caption {\label{tab:XQz5_sample} Properties of the quasars and their spectra in the XQz5 sample. The text contains a description of the table columns.} 
\begin{tabular}{lllrrrrrrrr}
\hline \hline
Name & RA (J2000) & Dec (J2000) & z$_{\rm{\mgii}}$$^{\rm{a}}$ & PID & E(B-V) & i$_{\rm{AB}}$$^{\rm{a}}$ & z$_{\rm{AB}}$$^{\rm{a}}$ & H$_{\rm{Vega}}$$^{\rm{a}}$ & SNR$_{1450}$ & SNR$_{2600}$ \\ 
\hline
SDSS J001115.23$+$144601.8 & 00:11:15.23 & 14:46:01.81 & 4.97 & X$_{\rm{3}}$\phantom{$,T_0$} & 0.045 & 18.28 & 18.10 & 16.53 & 122.8 & 63.5 \\ 
J001225$-$484830 & 00:12:25.02 & $-$48:48:30.10 & 4.62 & X$_{\rm{8}}$\phantom{$,T_0$} & 0.007 & 17.80 & 17.72 & 16.01 & 11.6 & 13.6 \\ 
SDSS J001714.67$-$100055.4 & 00:17:14.68 & $-$10:00:55.43 & 5.02 & X$_{\rm{5}}$\phantom{$,T_0$} & 0.034 & 19.57 & 19.62 & 17.57 & 14.8 & 12.2 \\ 
SDSS J002526.84$-$014532.5 & 00:25:26.84 & $-$01:45:32.52 & 5.06 & X$_{\rm{5}}$\phantom{$,T_0$} & 0.027 & 17.97 & 18.07 & 16.26 & 16.7 & 16.9 \\ 
SDSS J011546.27$-$025312.2 & 01:15:46.27 & $-$02:53:12.24 & 5.08 & X$_{\rm{5}}$\phantom{$,T_0$} & 0.031 & 20.05 & 19.82 & 17.98 & 10.9 & 4.5 \\ 
SDSS J013127.34$-$032100.1 & 01:31:27.35 & $-$03:21:00.07 & 5.20 & X$_{\rm{5}}$\phantom{$,T_0$} & 0.029 & 18.28 & 18.01 & 16.00 & 14.3 & 19.9 \\ 
J013539$-$212628 & 01:35:39.29 & $-$21:26:28.39 & 4.90 & X$_{\rm{8}}$\phantom{$,T_0$} & 0.015 & 17.97 & 17.82 & 15.84 & 14.0 & 23.0 \\ 
J014742$-$030248 & 01:47:41.53 & $-$03:02:47.88 & 4.80 & T$_{\rm{1}}$ & 0.021 & 18.42 & 18.00 & 16.29 & 4.7 & 12.9 \\ 
J015619$-$044140 & 01:56:18.99 & $-$04:41:39.84 & 4.93 & X$_{\rm{8}}$\phantom{$,T_0$} & 0.020 & 19.20 & 19.06 & 17.27 & 8.9 & 10.8 \\ 
J020437$-$252315 & 02:04:36.67 & $-$25:23:15.44 & 4.87 & X$_{\rm{8}}$,T$_{\rm{1}}$ & 0.012 & 18.29 & 18.35 & 16.59 & 6.2 & 10.2 \\ 
SDSS J021624.16$+$230409.4 & 02:16:24.16 & 23:04:09.47 & 5.23 & X$_{\rm{5}}$\phantom{$,T_0$} & 0.094 & 20.11 & 19.74 & 18.04 & 10.7 & 8.1 \\ 
SDSS J022112.62$-$034252.2 & 02:21:12.62 & $-$03:42:52.24 & 5.02 & X$_{\rm{5}}$\phantom{$,T_0$} & 0.019 & 19.42 & 19.59 & 17.75 & 11.5 & 5.5 \\ 
J022307$-$470902 & 02:23:06.76 & $-$47:09:02.73 & 5.00 & X$_{\rm{8}}$\phantom{$,T_0$} & 0.016 & 17.98 & 17.72 & 15.94 & 10.7 & 19.5 \\ 
J023649$-$114734 & 02:36:48.57 & $-$11:47:33.58 & 5.20 & X$_{\rm{9}}$,T$_{\rm{1}}$ & 0.021 & 18.80 & 18.37 & 16.84 & 10.8 & 15.9 \\ 
SDSS J024152.92$+$043553.4 & 02:41:52.92 & 04:35:53.46 & 5.19 & X$_{\rm{6}}$\phantom{$,T_0$} & 0.042 & 19.86 & 19.76 & 17.63 & 11.0 & 8.5 \\ 
SDSS J025121.33$+$033317.4 & 02:51:21.33 & 03:33:17.42 & 4.99 & X$_{\rm{5}}$\phantom{$,T_0$} & 0.058 & 19.29 & 19.31 & 17.35 & 14.7 & 15.9 \\ 
J033703$-$254832 & 03:37:03.06 & $-$25:48:31.55 & 5.11 & X$_{\rm{8}}$\phantom{$,T_0$} & 0.013 & 18.63 & 18.39 & 16.59 & 19.2 & 27.2 \\ 
SDSS J033829.30$+$002156.2 & 03:38:29.31 & 00:21:56.26 & 5.02 & X$_{\rm{5}}$\phantom{$,T_0$} & 0.077 & 20.10 & 19.90 & 18.19 & 14.3 & 11.5 \\ 
J040733$-$281031 & 04:07:32.97 & $-$28:10:31.26 & 4.74 & X$_{\rm{8}}$\phantom{$,T_0$} & 0.026 & 18.63 & 18.50 & 16.91 & 7.2 & 7.8 \\ 
J040915$-$275633 & 04:09:14.88 & $-$27:56:32.90 & 4.48 & T$_{\rm{1}}$ & 0.032 & 17.94 & 17.77 & 16.20 & 6.2 & 6.4 \\ 
J044433$-$292419 & 04:44:32.52 & $-$29:24:19.19 & 4.82 & T$_{\rm{1}}$ & 0.023 & 18.57 & 18.60 & 16.74 & 4.0 & 12.6 \\ 
J045057$-$265541 & 04:50:57.38 & $-$26:55:41.44 & 4.77 & X$_{\rm{8}}$\phantom{$,T_0$} & 0.031 & 18.74 & 18.48 & 16.43 & 9.6 & 12.0 \\ 
J045428$-$050049 & 04:54:27.95 & $-$05:00:49.46 & 4.83 & T$_{\rm{1}}$ & 0.027 & 18.46 & 18.46 & 16.56 & 5.7 & 16.7 \\ 
J051509$-$431854 & 05:15:08.93 & $-$43:18:53.64 & 4.61 & T$_{\rm{1}}$ & 0.022 & 18.52 & 18.35 & 16.25 & 6.5 & 9.2 \\ 
J072012$-$675632 & 07:20:11.67 & $-$67:56:31.56 & 4.62 & T$_{\rm{1}}$ & 0.117 & 18.39 & 18.09 & 16.22 & 1.9 & 18.9 \\ 
SDSS J074749.18$+$115352.4 & 07:47:49.17 & 11:53:52.48 & 5.26 & X$_{\rm{5}}$\phantom{$,T_0$} & 0.026 & 18.77 & 18.38 & 16.39 & 18.0 & 34.1 \\ 
SDSS J082454.01$+$130216.9 & 08:24:54.02 & 13:02:16.98 & 5.21 & X$_{\rm{4}}$\phantom{$,T_0$} & 0.030 & 20.03 & 19.55 & 17.89 & 23.5 & 17.0 \\ 
SDSS J084627.84$+$080051.7 & 08:46:27.84 & 08:00:51.73 & 5.02 & X$_{\rm{5}}$\phantom{$,T_0$} & 0.062 & 20.10 & 19.84 & 17.97 & 9.7 & 6.9 \\ 
SDSS J085430.37$+$205650.8 & 08:54:30.37 & 20:56:50.84 & 5.17 & X$_{\rm{5}}$\phantom{$,T_0$} & 0.023 & 19.42 & 19.75 & 17.66 & 11.3 & 7.8 \\ 
SDSS J090245.76$+$085115.9 & 09:02:45.76 & 08:51:15.92 & 5.22 & X$_{\rm{6}}$\phantom{$,T_0$} & 0.044 & 20.17 & 20.20 & 18.24 & 8.8 & 5.6 \\ 
J091656$-$251146 & 09:16:55.68 & $-$25:11:45.79 & 4.85 & X$_{\rm{8}}$,T$_{\rm{2}}$ & 0.118 & 17.33 & 17.28 & 15.45 & 14.1 & 25.0 \\ 
J093033$-$221208 & 09:30:32.57 & $-$22:12:07.75 & 4.89 & T$_{\rm{1}}$ & 0.058 & 18.22 & 18.23 & 16.49 & 5.6 & 25.4 \\ 
SDSS J095707.67$+$061059.5 & 09:57:07.67 & 06:10:59.52 & 5.17 & X$_{\rm{5}}$\phantom{$,T_0$} & 0.024 & 19.27 & 19.10 & 17.34 & 13.3 & 11.7 \\ 
SDSS J095712.20$+$101618.5 & 09:57:12.69 & 10:16:21.78 & 5.13 & X$_{\rm{6}}$\phantom{$,T_0$} & 0.036 & 20.22 & 20.11 & 18.23 & 4.8 & 4.8 \\ 
SDSS J095727.86$+$051905.2 & 09:57:27.93 & 05:19:05.70 & 5.19 & X$_{\rm{4}}$\phantom{$,T_0$} & 0.043 & 18.64 & 18.26 & 18.18 & 7.5 & 4.3 \\ 
SDSS J102833.45$+$074618.9 & 10:28:33.46 & 07:46:18.96 & 5.17 & X$_{\rm{4}}$\phantom{$,T_0$} & 0.022 & 20.25 & 19.97 & 18.38 & 8.2 & 5.7 \\ 
J111055$-$301130 & 11:10:54.69 & $-$30:11:29.94 & 4.78 & X$_{\rm{8}}$,T$_{\rm{2}}$ & 0.049 & 17.38 & 17.20 & 15.55 & 9.1 & 31.8 \\ 
J111520$-$193506 & 11:15:20.31 & $-$19:35:06.25 & 4.66 & X$_{\rm{8}}$,T$_{\rm{2}}$ & 0.049 & 18.40 & 18.36 & 16.75 & 23.0 & 17.0 \\ 
J113522$-$354839 & 11:35:21.99 & $-$35:48:38.84 & 4.97 & X$_{\rm{8}}$,T$_{\rm{2}}$ & 0.080 & 18.27 & 18.17 & 16.72 & 17.3 & 18.0 \\ 
SDSS J120055.61$+$181733.0 & 12:00:55.62 & 18:17:33.01 & 5.00 & X$_{\rm{5}}$\phantom{$,T_0$} & 0.021 & 19.71 & 19.60 & 17.97 & 16.5 & 13.8 \\ 
SDSS J120441.73$-$002149.6 & 12:04:41.73 & $-$00:21:49.54 & 5.09 & X$_{\rm{4}}$\phantom{$,T_0$} & 0.022 & 19.05 & 19.35 & 17.40 & 22.6 & 14.5 \\ 
BR 1202$-$0725 & 12:05:23.14 & $-$07:42:32.76 & 4.69 & X$_{\rm{2}}$\phantom{$,T_0$}& 0.032 & 17.94 & 17.85 & 16.22 & 59.9 & 71.4 \\ 
SDSS J12083$+$0010 & 12:08:23.83 & 00:10:27.60 & 5.28 & X$_{\rm{1}}$\phantom{$,T_0$} & 0.020 & 20.68 & 20.36 & 18.82 & 32.9 & 18.3 \\ 
J121402$-$123548 & 12:14:02.71 & $-$12:35:48.75 & 4.75 & T$_{\rm{2}}$ & 0.041 & 18.61 & 18.43 & 16.80 & 4.0 & 17.8 \\ 
J121921$-$360933 & 12:19:21.12 & $-$36:09:33.13 & 4.80 & X$_{\rm{8}}$\phantom{$,T_0$} & 0.068 & 18.96 & 18.52 & 16.51 & 8.0 & 22.3 \\ 
\hline \hline
\multicolumn{10}{l}{\footnotesize
$^{\rm{a}}$Uncertainties in the full table available as supplementary material.}\\
\multicolumn{11}{l}{\footnotesize
X$_{\rm{1}}$ -- 084.A-0574(A); X$_{\rm{2}}$ -- 084.A-0780(B); X$_{\rm{3}}$ -- 087.A-0125(A); X$_{\rm{4}}$ -- 094.A-0793(A); X$_{\rm{5}}$ -- 098.A-0111(A); X$_{\rm{6}}$ -- 0100.A-0243(A);}\\
\multicolumn{11}{l}{\footnotesize
X$_{\rm{7}}$ -- 0104.A-0410(A); X$_{\rm{8}}$ -- 108.22H9.001; X$_{\rm{9}}$ -- 109.23D1.001; X$_{\rm{10}}$ -- 109.23D1.002; T$_{\rm{1}}$ -- 2021B-0036; T$_{\rm{2}}$ -- 2022A-389756}
\end{tabular}
\end{table*}
\endgroup
\setcounter{table}{0}
\begingroup
\begin{table*}
\caption {(Continued)}  
\begin{tabular}{lllrrrrrrrr}
\hline \hline
Name & RA (J2000) & Dec (J2000) & z$_{\rm{\mgii}}$ & PID & E(B-V) & i$_{\rm{AB}}$ & z$_{\rm{AB}}$ & H$_{\rm{Vega}}$ & SNR$_{1450}$ & SNR$_{2600}$ \\ 
\hline
J130031$-$282931 & 13:00:31.13 & $-$28:29:30.99 & 4.71 & T$_{\rm{1}}$ & 0.069 & 18.34 & 18.03 & 16.35 & 4.2 & 20.1 \\ 
SDSS J133257.45$+$220835.9 & 13:32:57.44 & 22:08:35.86 & 5.11 & X$_{\rm{6}}$\phantom{$,T_0$} & 0.010 & 19.22 & 19.29 & 17.33 & 14.4 & 9.3 \\ 
J140802$-$275820 & 14:08:01.82 & $-$27:58:20.36 & 4.47 & T$_{\rm{2}}$ & 0.043 & 17.83 & 17.67 & 16.05 & 11.7 & 11.7 \\ 
SDSS J142325.92$+$130300.7 & 14:23:25.92 & 13:03:00.70 & 5.04 & X$_{\rm{5}}$\phantom{$,T_0$} & 0.019 & 19.56 & 19.39 & 17.62 & 18.2 & 16.7 \\ 
J142721$-$050353 & 14:27:21.56 & $-$05:03:53.04 & 5.09 & X$_{\rm{9}}$\phantom{$,T_0$} & 0.051 & 19.24 & 18.87 & 17.21 & 6.0 & 11.4 \\ 
SDSS J143605.00$+$213239.2 & 14:36:05.00 & 21:32:39.25 & 5.23 & X$_{\rm{5}}$\phantom{$,T_0$} & 0.024 & 20.08 & 19.70 & 17.95 & 14.8 & 9.9 \\ 
J151443$-$325024 & 15:14:43.82 & $-$32:50:24.92 & 4.83 & X$_{\rm{10}}$,T$_{\rm{2}}$ & 0.300 & 18.01 & 17.83 & 15.82 & 5.0 & 11.0 \\ 
J153241$-$193033 & 15:32:41.40 & $-$19:30:32.79 & 4.69 & X$_{\rm{9}}$,T$_{\rm{2}}$ & 0.111 & 19.03 & 18.80 & 16.65 & 10.7 & 12.2 \\ 
J153359$-$181027 & 15:33:59.76 & $-$18:10:27.20 & 5.01 & X$_{\rm{9}}$\phantom{$,T_0$} & 0.091 & 19.22 & 19.04 & 16.91 & 4.5 & 11.7 \\ 
J155657$-$172107 & 15:56:57.36 & $-$17:21:07.51 & 4.75 & X$_{\rm{9}}$,T$_{\rm{2}}$ & 0.175 & 18.60 & 18.46 & 16.66 & 5.8 & 19.6 \\ 
SDSS J160111.16$-$182835.0 & 16:01:11.17 & $-$18:28:35.09 & 5.05 & X$_{\rm{5}}$\phantom{$,T_0$} & 0.363 & 19.98 & 19.46 & 17.34 & 12.5 & 21.2 \\ 
SDSS J161622.10$+$050127.7 & 16:16:22.11 & 05:01:27.72 & 4.87 & X$_{\rm{3}}$\phantom{$,T_0$} & 0.059 & 18.93 & 18.78 & 17.24 & 114.9 & 106.8 \\ 
J162551.54$-$043049.4 & 16:25:51.55 & $-$04:30:49.52 & 5.20 & T$_{\rm{1}}$,T$_{\rm{2}}$ & 0.186 & 18.69 & 18.40 & 16.54 & 2.7 & 15.5 \\ 
J194124$-$450023 & 19:41:24.59 & $-$45:00:23.77 & 5.21 & X$_{\rm{9}}$,T$_{\rm{2}}$ & 0.067 & 18.43 & 18.19 & 16.45 & 11.6 & 17.6 \\ 
J205559$-$601147 & 20:55:59.23 & $-$60:11:47.42 & 4.97 & X$_{\rm{9}}$\phantom{$,T_0$} & 0.045 & 18.78 & 18.87 & 17.22 & 15.5 & 19.3 \\ 
J211105$-$015604 & 21:11:05.61 & $-$01:56:04.21 & 4.89 & X$_{\rm{9}}$\phantom{$,T_0$} & 0.043 & 18.00 & 17.88 & 16.39 & 12.6 & 15.6 \\ 
J211921$-$772253 & 21:19:20.86 & $-$77:22:53.27 & 4.56 & X$_{\rm{10}}$,T$_{\rm{2}}$ & 0.093 & 17.96 & 17.61 & 16.03 & 7.4 & 4.6 \\ 
J214608$-$485819 & 21:46:08.22 & $-$48:58:19.59 & 5.16 & X$_{\rm{9}}$,T$_{\rm{2}}$ & 0.023 & 18.39 & 18.28 & 16.44 & 10.7 & 12.4 \\ 
J215728.21$-$360215.1 & 21:57:28.23 & $-$36:02:15.21 & 4.70 & X$_{\rm{7}}$\phantom{$,T_0$} & 0.015 & 17.26 & 17.07 & 14.80 & 52.0 & 83.7 \\ 
SDSS J220106.63$+$030207.7 & 22:01:06.63 & 03:02:07.67 & 5.09 & X$_{\rm{5}}$\phantom{$,T_0$} & 0.039 & 19.37 & 19.24 & 17.30 & 23.8 & 17.1 \\ 
J220159$-$202627 & 22:01:58.60 & $-$20:26:27.38 & 4.74 & X$_{\rm{9}}$\phantom{$,T_0$} & 0.026 & 18.49 & 18.13 & 16.42 & 12.5 & 9.6 \\ 
SDSS J220226.77$+$150952.3 & 22:02:26.77 & 15:09:52.38 & 5.08 & X$_{\rm{5}}$\phantom{$,T_0$} & 0.041 & 18.89 & 18.68 & 17.10 & 17.7 & 8.4 \\ 
J221112$-$330246 & 22:11:11.55 & $-$33:02:45.91 & 4.65 & X$_{\rm{8}}$\phantom{$,T_0$} & 0.018 & 18.35 & 18.08 & 15.94 & 11.7 & 17.6 \\ 
SDSS J221644.01$+$001348.1 & 22:16:44.02 & 00:13:48.12 & 5.01 & X$_{\rm{4}}$\phantom{$,T_0$} & 0.052 & 20.37 & 20.29 & 18.60 & 15.3 & 8.7 \\ 
J222153$-$182603 & 22:21:52.88 & $-$18:26:02.93 & 4.53 & T$_{\rm{1}}$ & 0.033 & 17.63 & 17.36 & 15.78 & 17.1 & 17.6 \\ 
J222358$-$252634 & 22:23:57.88 & $-$25:26:34.40 & 4.80 & X$_{\rm{8}}$\phantom{$,T_0$} & 0.020 & 18.78 & 18.72 & 16.96 & 6.7 & 7.6 \\ 
SDSS J222514.38$+$033012.5 & 22:25:14.38 & 03:30:12.50 & 5.26 & X$_{\rm{5}}$\phantom{$,T_0$} & 0.086 & 18.65 & 18.25 & 17.99 & 9.8 & 15.5 \\ 
J222612$-$061807 & 22:26:12.42 & $-$06:18:07.37 & 5.10 & X$_{\rm{8}}$\phantom{$,T_0$} & 0.052 & 18.67 & 18.78 & 17.09 & 9.9 & 9.2 \\ 
SDSS J222845.14$-$075755.3 & 22:28:45.15 & $-$07:57:55.38 & 5.16 & X$_{\rm{6}}$\phantom{$,T_0$} & 0.042 & 20.05 & 20.06 & 17.94 & 13.5 & 10.4 \\ 
J223419$-$804013 & 22:34:19.12 & $-$80:40:13.30 & 4.97 & X$_{\rm{9}}$\phantom{$,T_0$} & 0.111 & 18.74 & 18.70 & 17.22 & 11.9 & 7.1 \\ 
J230349$-$063343 & 23:03:49.20 & $-$06:33:43.18 & 4.74 & X$_{\rm{10}}$,T$_{\rm{1}}$ & 0.037 & 18.26 & 17.83 & 15.96 & 4.8 & 13.7 \\ 
J230430$-$313427 & 23:04:29.89 & $-$31:34:27.11 & 4.87 & T$_{\rm{1}}$ & 0.022 & 17.73 & 17.73 & 16.13 & 5.4 & 17.0 \\ 
SDSS J232536.64$-$055328.3 & 23:25:36.64 & $-$05:53:28.43 & 5.23 & X$_{\rm{6}}$\phantom{$,T_0$} & 0.030 & 19.75 & 19.10 & 17.68 & 10.7 & 7.5 \\ 
J232953$-$200039 & 23:29:52.78 & $-$20:00:39.19 & 5.04 & X$_{\rm{9}}$\phantom{$,T_0$} & 0.028 & 18.52 & 18.39 & 16.55 & 11.4 & 14.4 \\ 
J233435$-$365709 & 23:34:35.30 & $-$36:57:08.99 & 4.72 & X$_{\rm{9}}$\phantom{$,T_0$} & 0.014 & 19.27 & 18.90 & 16.51 & 9.3 & 16.7 \\ 
J233505$-$590103 & 23:35:05.90 & $-$59:01:03.39 & 4.53 & X$_{\rm{7}}$\phantom{$,T_0$} & 0.012 & 17.58 & 17.57 & 15.92 & 60.1 & 45.4 \\ 
SDSS J234433.50$+$165316.4 & 23:44:33.50 & 16:53:16.56 & 4.99 & X$_{\rm{5}}$\phantom{$,T_0$} & 0.052 & 18.65 & 18.74 & 16.86 & 6.6 & 5.8 \\ 
SDSS J235124.31$-$045907.3 & 23:51:24.31 & $-$04:59:07.30 & 5.25 & X$_{\rm{6}}$\phantom{$,T_0$} & 0.030 & 20.41 & 20.07 & 18.36 & 10.8 & 7.0 \\ 

\hline \hline
\end{tabular}
\end{table*}
\endgroup

\begingroup
\begin{table}
\centering{
\caption {\label{tab:XQz5_line_properties} Description of measured properties from modelling the spectral 
continuum and \mgii\ broad emission-line.}  
\begin{tabular}{lll}
\hline \hline
 Property & Description  & Units\\
 \hline
Redshift & Defined by the peak flux of the line model & \\
FWHM & Full-width half-maximum of profile & km s$^{-1}$\\
Sigma & Second moment of profile & km s$^{-1}$ \\
Blueshift & Defined by the flux-bisecting wavelength &  km s$^{-1}$ \\
EW & Equivalent width in rest-frame & \AA \\
$\log$ iLuminosity & Integrated log line luminosity & erg s$^{-1}$ \\
$\log \rm{L}_{\rm{3000}}$ & Power-law continuum luminosity at 3000\AA\ & erg s$^{-1}$ \\
\hline \hline
\end{tabular}
}
\end{table}
\endgroup

\section{Spectral Modelling} \label{sec:spectral_fitting}
We measure the properties of the \mgiil\ emission feature 
to constrain the redshifts of our target quasars and determine black hole masses from the velocity broadening. 
Here, we describe our emission-line modelling approach, which utilises a publicly available code \citep[\texttt{PyQSpecFit}\footnote{\hyperlink{https://github.com/samlaihei/PyQSpecFit}{https://github.com/samlaihei/PyQSpecFit}};][]{PyQSpecFit_v1} designed 
for modelling quasar spectral lines,
%
including the components of 
our continuum and emission-line models. More information on the code and each of these components can be found in \citet{Lai23_XQ100}, which is a study of the black hole properties in the XQ-100 quasar legacy survey.

\begin{figure*}
	\includegraphics[width=0.9\textwidth]{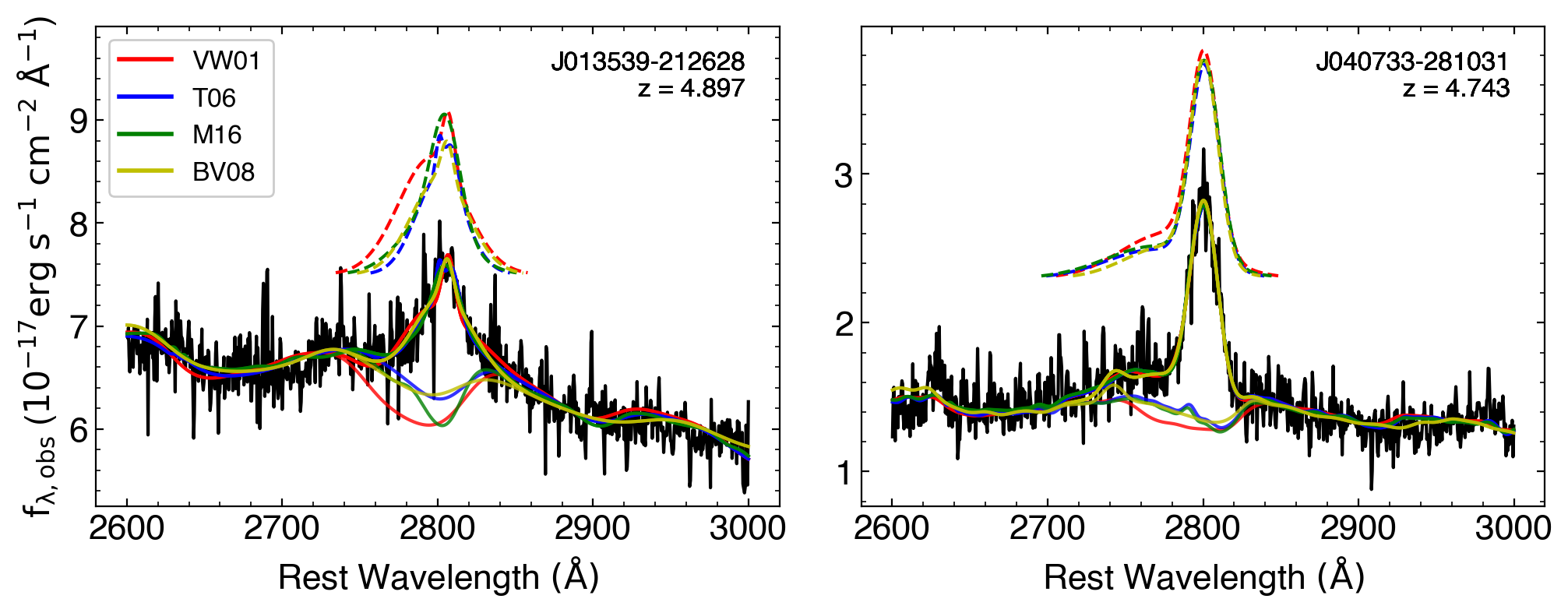}
    \caption{Example models of the \mgii\ line for J013539--212628 (left) and J040733--281031 (right), where the adopted underlying \feii\ continuum model is varied between the four templates discussed in Section \ref{sec:cont-fitting}. The continuum and line models between templates are differentiated by colour and the continuum-subtracted broad line model is represented with the dashed lines, vertically shifted for visibility. The targets in each panel show the wide range in responsiveness of the \mgii\ profile to the choice of \feii\ template: the target on the left panel shows a highly sensitive line model whereas the line model on the right is robust.} \label{fig:varied_FeII}
\end{figure*}

\subsection{Continuum Modelling} \label{sec:cont-fitting}
Before we model the \mgii\ line, we first subtract a model of the quasar continuum. Our quasar continuum model is composed of two components: a power law and \feii\ template, which is altogether called the pseudo-continuum. The pseudo-continuum is then uniquely defined by five free parameters and is fit simultaneously to the spectroscopic data contained within selected rest-frame windows: 1973--1983\AA, 2060--2340\AA, 2600--2740\AA, and 2840--3100\AA, chosen to minimise contribution from strong emission-lines while also covering wavelength regions in close proximity to the \mgii\ line with a wide diversity of expected \feii\ flux. 

The \mgii\ feature is sensitive to the \feii\ continuum emission, which extends underneath the emission-line, and it has previously been established that the choice of \feii\ template can bias the result. Notably, adopting the \citet{Vestergaard_2001} template, which is often used to calibrate the \mgii-based single-epoch virial black hole mass estimate \citep[e.g.,][]{Vestergaard_2009, Shen_2011}, results in larger line widths than with other models of the \feii\ emission \citep{Schindler_2020, Lai23_XQ100}. Thus, we rely on four empirical and semi-empirical \feii\ templates: \citet[][VW01]{Vestergaard_2001}, \citet[][T06]{Tsuzuki_2006}, \citet[][BV08]{Bruhweiler_Verner_2008}, and \citet[][M16]{Mejia-Restrepo_2016} to minimise bias induced by any single template. The templates are renormalised and broadened with a Gaussian kernel to match the features of each observed spectrum. 

A third potential component to the pseudo-continuum, the Balmer continuum, is sometimes introduced in similar studies of the \mgii\ emission-line in AGN \citep[e.g.,][]{Wang_2009}. However, the flux of the Balmer continuum decreases continuously on the blue side of the Balmer edge located at 3645\AA, and is relatively weak near the \mgii\ line \citep{Vestergaard_2001, Dietrich_2002}. We also note that the Balmer continuum flux can be degenerate with the other pseudo-continuum components. Thus, despite an improved global continuum fit, the isolation of the \mgii\ emission-line profile is often independent of a reasonably constrained Balmer continuum model \citep[e.g.,][]{Shen_2012, Lai23_XQ100}. Therefore, we do not include the Balmer continuum in our pseudo-continuum model, although we note that the exclusion of the Balmer continuum can change the flux of the power-law and \feii\ components. 

Even though we do not observe evidence of a strong Balmer flux contribution across the Balmer edge in the luminous quasar spectra of XQ-100 \citep{Lai23_XQ100}, we estimate the systematic uncertainty related to the exclusion of the Balmer contribution to the continuum model. A common model of the Balmer continuum, adopted in other studies \citep[e.g.][]{Dietrich_2003, Kurk_2007, DeRosa_2011, Shin_2019}, normalises the flux at the Balmer edge to 30\% the flux of the power-law component at 3675\AA, where the \feii\ emission is absent. We assume a lower normalisation of 10\% that is also commonly used in the literature \citep{Calderone_2017_QSFit, Rakshit_2020}. Following these studies, we adopt the same electron temperature ($T_{\rm{e}}$ = 15,000 K) and optical depth ($\tau_{\lambda}$ = 1), finding that the Balmer flux contribution at 3000\AA\ would be between 7--8\% of the power-law flux, depending on the power-law slope, where we have used the 16$^{\rm th}$ and 84$^{\rm th}$ percentiles of all measured slopes from our sample, $\gamma = -1.36^{+0.40}_{-0.33}$. Thus, the power-law flux at 3000\AA\ could be overestimated by up to 10\% depending on how the Balmer flux is proportioned between the power-law and \feii\ components. We later quantify on how this could affect our results.

\subsection{Mg\,\textsc{ii} Modelling} \label{sec:line-fitting}
Broad emission-line profiles can be found with a wide range of properties and morphological complexities, which render emission-line models described by a single analytical function unsuitable, particularly with a Doppler broadened doublet emission-line like \mgii. As such, quasar spectral modelling studies typically utilise a multiple Gaussian approach to fit each individual emission feature \citep[e.g.,][]{Greene_2005, Shen_2011, Rakshit_2020, Wu_2022}. Accordingly, we model the \mgii\ broad emission-line with three Gaussian components and use an additional component to fit the narrow line, if present. The maximum width of the narrow component is set at 1000 km s$^{-1}$. 

After subtracting the pseudo-continuum model, we fit the \mgii\ line model within rest-frame 2730--2870\AA. We then determine the \mgii\ emission-line properties from the total line profile composed of the three broad components. The final mean line properties and measurement uncertainties are determined from the emergent emission-line models assuming the four different \feii\ templates. We further add in quadrature the additional uncertainty measured by fitting 50 synthetic spectra, which are generated by randomly redistributing the flux in each fixed-velocity bin according to its Gaussian uncertainties.

In Table \ref{tab:XQz5_line_properties}, we describe each of the line properties measured from the quasar continuum and \mgii\ broad emission-line model. For each target, we update the measured redshift based on the wavelength of the peak flux in the reconstructed \mgii\ line profile. From the combined broad emission-line profile, we measure the \mgii\ FWHM and line dispersion, Sigma, which is the second moment of the line profile. The blueshift of the line is measured from the flux-bisecting wavelength, as
\begin{equation} \label{eq:blueshift}
    \frac{\rm{Blueshift}}{\rm{km \,s^{-1}}} \equiv c \times (\lambda_{\rm{ref}}- \lambda_{\rm{med}})/\lambda_{\rm{ref}}\,,
\end{equation}
where $\lambda_{\rm{ref}}$ is the vacuum wavelength of the \mgii\ doublet which is set to 2799.117\AA, and $\lambda_{\rm{med}}$ is the wavelength bisecting the total continuum-subtracted emission-line flux. We also measure the line rest-frame equivalent width (EW) and integrated luminosity. The monochromatic luminosity at 3000\AA\ is measured from the power-law continuum model. For the 14 targets with more than one spectrum, the final line properties are measured from the weighted average of multiple observations. In our sample, we find one target, J051509--431854, with clear \mgii\ absorption, which is a common characteristic of low-ionisation broad absorption line (LoBAL) quasars. This is consistent with the expectation that quasar with BAL outflows are found in less than 20\% of quasars \citep[e.g.][]{Gibson_2009} and LoBAL outflows are even rarer, consisting of 10--15\% of BAL quasars. However, there is some evidence that the BAL fraction increases with redshift \citep{Bischetti_2023}. A more detailed analysis of the BAL quasar fraction will be in an separate forthcoming study.

In Figure \ref{fig:varied_FeII}, we show two targets which represent a range of line model susceptibility to the \feii\ template. The target on the left panel (J013539--212628) is highly sensitive to the \feii\ model, with a \mgii\ FWHM that varies from 2910--5400 km s$^{-1}$. In contrast, the target on the right panel (J040733--281031) is minimally affected as a result of its weak \feii\ emission. Its \mgii\ FWHM range is from 2420--2540 km s$^{-1}$.

\subsubsection{Single-Epoch Virial Mass Estimate} \label{sec:virial_mbh}
We use the version of the \mgii\ single-epoch virial mass estimate from \citet{Shen_2011}, which is calibrated to a high-luminosity subset from the local reverberation mapping AGN sample using the \hbetal\ line,
\begin{equation}
   \left(\frac{\rm M_{\rm{BH,vir}}}{\rm M_{\odot}}\right) = 10^{6.74} \left[\frac{\rm L(3000)}{10^{44} \,\rm{erg\, s^{-1}}}\right]^{0.62} \left[\frac{\rm{FWHM_{\mgii}}}{1000 \,\rm{km\, s^{-1}}}\right]^{2} \,,
   \label{eq:mgii_virial}
\end{equation}
where $L(3000)$ is the monochromatic luminosity ($\lambda L_{\lambda}$) of the quasar continuum at 3000\AA\ and FWHM$_{\rm{\mgii}}$ is the measured line full-width at half-maximum of the \mgii\ broad line profile. Systematic errors from the single-epoch virial black hole mass estimate can be as high as 0.5 dex due to the combined effects of a 0.3 dex scatter around their reverberation mapping counterparts \citep{DallaBonta_2020} and a 0.4 dex intrinsic scatter of reverberation-based estimates around the $M_{\rm BH}-\sigma_{*}$ relation \citep{Bennert_2021}. We adopt 0.5 dex as our single-epoch virial black hole mass uncertainty in this study. A possible 25\% overestimation in $\rm L(3000)$ as suggested in Section \ref{sec:cont-fitting} would overestimate the black hole mass by 0.075 dex, which is insignificant compared to the systematic uncertainties of the method.

Although we measure the line dispersion, we opt to use the FWHM in the determination of black hole mass. The line dispersion may have advantages over the FWHM \citep[e.g.,][]{Fromerth_2000, Peterson_2004, Collin_2006}, but in practice, it is sensitive to the wings of the line profile, which are naturally low in flux and difficult to disentangle from the pseudo-continuum model. For the XQz5 sample, we find that the fractional uncertainty for the line dispersion is typically higher than that of the FWHM. 

With the single-epoch virial mass estimate, the typical black hole mass measurement uncertainty in our sample is 0.11 dex, based on statistical uncertainties in the measured FWHM and $\rm L(3000)$. The error in both of these quantities is composed of the measured variance from adopting different \feii\ models and fitting synthetic spectra. 

\begin{figure*}
	\includegraphics[width=0.9\textwidth]{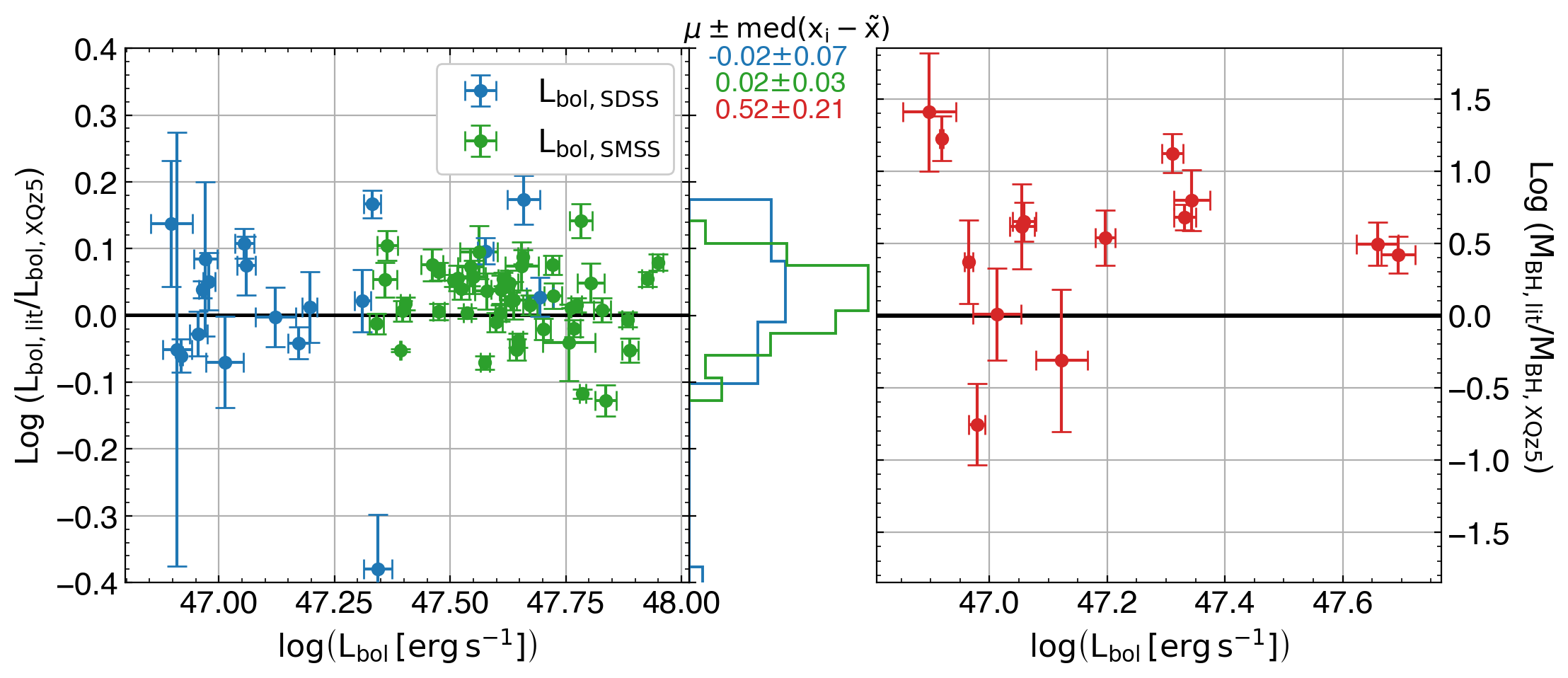}
    \caption{Comparisons of estimated bolometric luminosity (left) and black hole masses (right) between XQz5 measurements and the existing literature for a subset of the XQz5 with SDSS DR16Q measurements \citep{Wu_2022} or \citetalias{Onken_2022_QLF} luminosities. The SDSS bolometric luminosities are estimated with the bolometric correction factor $k_{\rm{1350}} = 3.81$, while the SMSS and XQz5 luminosities are estimated with $k_{\rm{3000}} = 5.15$. The black hole masses from SDSS are measured from \civ. The mean and median absolute deviation of the residual for each measured quantity is listed There is a tight agreement with literature bolometric luminosities, but the \civ-based virial mass estimates from SDSS are overestimated compared to XQz5 \mgii-based $M_{\rm{BH}}$ measurements.}
    \label{fig:Lit_comp}
\end{figure*}

\begin{figure}
	\includegraphics[width=1.0\columnwidth]{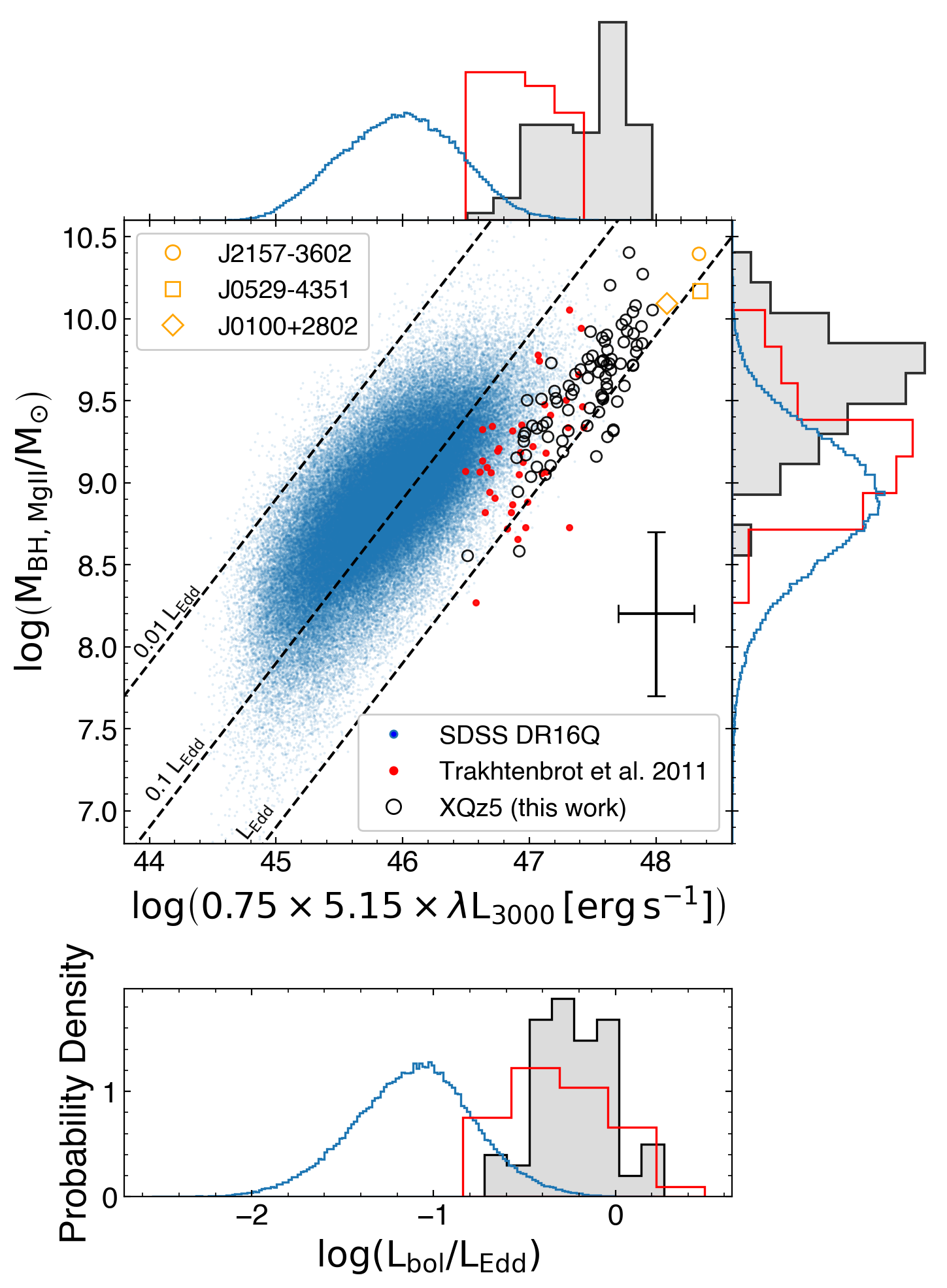}
    \caption{Black hole mass and bolometric luminosity density distributions as measured from k-corrections to the observed 3000\AA\ monochromatic luminosity. Our sample, XQz5 (black open circles and shaded histograms), is compared to the SDSS DR16Q sample \citep[blue points and histograms;][]{Wu_2022} with \mgii\ measurements (at z = 0.35 -- 2.3) and the \citetalias{Trakhtenbrot_2011} sample (red points and histograms), which is at a comparable redshift. By construction, XQz5 is the highest luminosity sample at this redshift range and we found it to be skewed towards higher black hole masses. However, the Eddington ratio distribution is not statistically different from the \citetalias{Trakhtenbrot_2011} sample. We assume typical systematic errors of $\Delta \log{M} = 0.5$ and $\Delta \log{L_{\rm{bol}}} = 0.3$, which is shown with the cross. We have also reproduced the three notable quasars presented in Figure \ref{fig:sample_comparison} (open orange symbols in the $\rm{M_{BH}}-\rm{L_{bol}}$ panel).}
    \label{fig:Mbh-Lbol}
\end{figure}

\section{Results and Discussion} \label{sec:results_discussion}

The results for the parameters given in Table \ref{tab:XQz5_line_properties} are provided in a supplementary file, which is also available in a public repository\footnote{\href{https://github.com/samlaihei/XQz5}{https://github.com/samlaihei/XQz5}} and a sample of the measured properties is shown in Table \ref{tab:MgII_props}. We present fits of the \mgii\ line for all 83 quasars in Appendix \ref{appendix:additional_tables_figs}. Although some targets from the ESO Science Archive Facility have existing \mgii-based black hole mass estimates, such as SDSS J013127.34--032100.1 \citep{Yi_2014} and SDSS J161622.10$+$050127.7 \citepalias{Trakhtenbrot_2011}, we remeasure the black hole mass in a homogeneous fashion and place them in context within the larger sample. The estimated black hole mass range of the XQz5 sample is $\log{(\rm{M_{BH}}/\rm{M_\odot})} = 8.5-10.4$.

In this section, we discuss the corrections used to estimate bolometric luminosity from the observed spectrum and its associated uncertainties. We then compare the derived quantities between quasar samples collected at similar redshifts, and luminosity-matched samples at different redshifts. We also compare the measured properties to other high-quality X-shooter samples, XQ-100 and XQR-30.

\begingroup
\begin{table*}
\caption {\label{tab:MgII_props} Examples of the measured properties from the model of the quasar continuum and \mgii\ line. Table \ref{tab:XQz5_line_properties} presents a description of the columns. The full tabulated results are provided in a supplementary file.}  
\begin{tabular}{lccccccc}
\hline \hline
ID & FWHM & Sigma & Blueshift & EW &  $\log{\rm{iLuminosity}}$ & $\log{\rm{L(3000)}}$ & $\log{\rm{M_{BH}}}$ \\ 
\hline
J040733-281031 & 2485 $\pm$ \phantom{0}92 & 2515 $\pm$ 232 & 205 $\pm$ \phantom{0}57 & 37.66 $\pm$ 2.63 & 44.80 $\pm$ 0.03 & 46.67 $\pm$ 0.01 & 9.19 $\pm$ 0.03 \\ 
J091656-251146 & 3057 $\pm$ 174 & 1793 $\pm$ 133 & --62 $\pm$ 101 & 15.61 $\pm$ 2.02 & 45.09 $\pm$ 0.05 & 47.32 $\pm$ 0.01 & 9.77 $\pm$ 0.05 \\ 
J211105-015604 & 3821 $\pm$ 378 & 2392 $\pm$ 193 & 130 $\pm$  \phantom{0}92 & 24.73 $\pm$ 2.13 & 45.11 $\pm$ 0.03 & 47.15 $\pm$ 0.01 & 9.86 $\pm$ 0.09 \\ 
\hline \hline
\end{tabular}
\end{table*}
\endgroup

\subsection{Bolometric Luminosity} 

To measure the bolometric luminosity of a quasar exactly, it is necessary to observe the quasar at all wavelengths from all possible viewing angles and have an understanding of the source component decomposition for the observed luminosity at different wavelengths. Because that is not feasible, it is common among studies of quasars to apply a first-order correction to the observed monochromatic luminosity in the form, $\rm L_{\rm{bol}} = k_{\lambda}\rm L(\lambda)$, where $k_{\lambda}$ is the bolometric correction factor and $\rm L(\lambda)$ represents the monochromatic luminosity, $\lambda \rm L_{\lambda}$ at rest wavelength $\lambda$.

The bolometric correction factor, $k_{\lambda}$, is calibrated for a mean quasar spectral energy distribution (SED), integrated across a particular wavelength range. In this study, we use $k_{3000} = 5.15$, a first-order correction from \citet{Richards_2006} that is calibrated to the mean SED of 259 SDSS Type 1 quasars, integrated from 100 $\mu$m to 10 keV and assumed to be emitting isotropically. However, as the Type 1 quasars used to construct the mean SED are likely to be biased towards face-on orientations, the resulting luminosity will be overestimated. Thus, we adopt the model presented in \citet{Lai_23_AD} and assume that the maximum opening angle of the obscuring torus for our sample, which defines the boundary between a Type 1 and Type 2 quasar classification, is 65$^{\circ}$. We then use a thin disc model \citep{Li_2005} to estimate the correction between the observed and bolometric luminosities, assuming the quasars are randomly distributed in observed orientations and black hole spins. Using this method, the mean correction factor for anisotropy is calculated to be a 25\% luminosity suppression to the observed luminosity, which is consistent with the correction in \citet{Runnoe_2012}, derived from similar arguments. Thus, for this study, we adopt $\rm L_{\rm{bol}} = 0.75 \times 5.15 \times \rm L(3000)$ and we use this bolometric correction consistently for the comparison with literature samples, unless otherwise stated. Due to the spectral diversity of quasars, the systematic uncertainty from deriving the assuming a mean SED can be as high as 50\% \citep{Richards_2006}, and so a systematic error of $\Delta \log{\rm L_{\rm{bol}}} = 0.3$ is assumed. We further note that the bolometric correction is expected to be mass or luminosity-dependent \citep[e.g.][]{Runnoe_2012, Netzer_2019}. For an ultraluminous 24 billion solar mass black hole, the bolometric correction factor at 3000\AA\ could be as low as $k_{3000} = 1.62$ \citep{Lai_23_AD}, which is less than half of the bolometric luminosity derived with this method. An additional 0.05 dex adjustment may also result from a 10\% overestimation of continuum flux by not explicitly modelling the Balmer component as cautioned in Section \ref{sec:cont-fitting} and calibration errors may contribute up to 0.1 dex, but these corrections are not as significant as the assumed systematic error. We emphasise the possibility that the brightest and most massive black holes in this sample could have overestimated bolometric luminosities, but we adopt a consistent correction for all targets to facilitate comparison.

Using $k_{3000} = 5.15$ with a 0.75 anisotropy correction factor, the estimated bolometric luminosity range occupied by the XQz5 sample is then $\log{(\rm{L_{bol}}/\rm{erg\,s^{-1}})} = 46.5-48.34$ with Eddington ratios spanning nearly an order of magnitude from 0.19--1.8.

\subsection{Comparison with literature samples}

\begin{figure*}
	\includegraphics[width=0.9\textwidth]{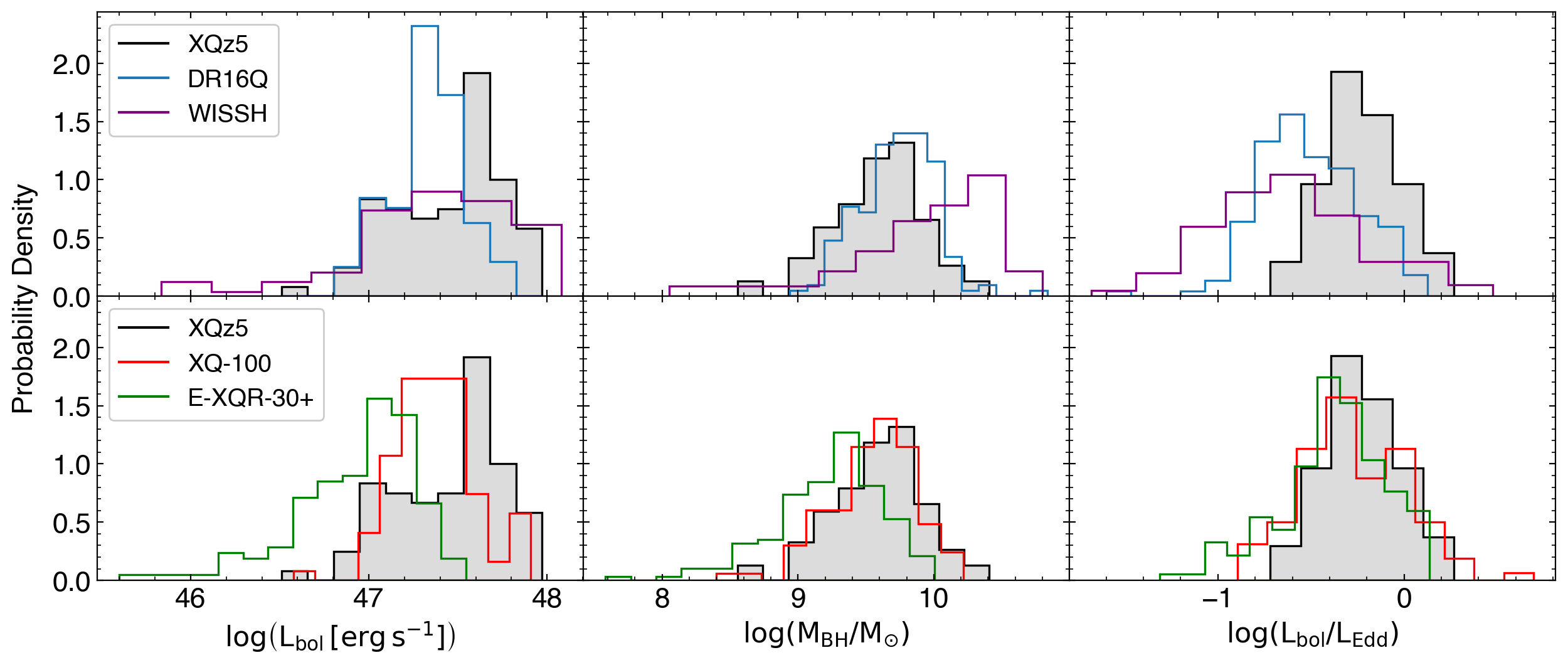}
    \caption{Bolometric luminosity (left), black hole mass (middle), and Eddington ratio (right) distributions for XQz5 (shaded) in comparison to two lower-redshift samples (SDSS DR16Q and WISSH) of similar luminosity (top) and two other brightness-selected quasar samples (XQ-100 and E-XQR-30+) with NIR spectroscopic data at lower and higher redshifts (bottom). The version of DR16Q plotted here (in blue) is a luminosity-matched subset of SDSS DR16Q. WISSH (in purple) is a high-luminosity SDSS and WISE-selected sample designed to reflect the most luminous known AGN in the universe. Both XQ-100 and E-XQR-30+ are bright quasar samples with high-quality spectra from X-shooter.} \label{fig:Comp_hists}
\end{figure*}

In this section, we place our black hole mass, luminosity, and Eddington ratio estimates in context with the literature. First, we compare the measurements obtained from the SDSS quasars in our sample with the results from the \civ-based virial estimator from SDSS DR16Q \citep{Wu_2022}. We also compare our estimates of bolometric luminosity with those from \citetalias{Onken_2022_QLF}\footnote{\citetalias{Onken_2022_QLF} does not include black hole mass estimates}, which derived its $L_{\rm{3000}}$ luminosity estimates by matching each target's photometric information to a composite spectrum of bright quasars from $z=1-2$ \citep{Selsing_2016}. These comparisons are presented in Figure \ref{fig:Lit_comp}, from which we observe that the bolometric luminosity estimates between XQz5 and the literature are in tight agreement, despite the fact that the DR16Q estimates come from a different wavelength regime as $\rm L_{\rm{bol}} = 3.81\,\rm L_{\rm{1350}}$. The literature luminosity of one outlier (not shown), identified as SDSS J001714.67--100055.4, is 0.7 dex lower than our measured luminosity and we found its continuum model in DR16Q to have been affected by a broad absorption feature. As for the measured black hole masses, the \civ-based virial mass estimator applied to SDSS data significantly overestimates the $M_{\rm{BH}}$ for our sample on average. When the 1690--1705\AA\ continuum fitting window falls near the red end of the wavelength coverage affected by enhanced sky background and molecular oxygen skylines, there is no anchor point for the continuum model on the red side of \civ. For this overlapping sample, we find that the cubic polynomial continuum model component adopted in \citep{Wu_2022} significantly underestimates the continuum underneath the \civ\ line when fitted only to the blue side of \civ, which results in inflated \civ\ FWHMs and overestimated black hole masses.

Before comparing the XQz5 to other quasar samples, we first caution that systematic biases can arise from heterogeneous subsample selection criteria, as even well-defined subsamples can be biased if the sample completeness is unknown \citep{Shen_2012_Demo, Kelly_2013, Wu_2022_Demo}. Furthermore, we have defined the bolometric luminosity estimate using a static correction to the UV continuum luminosity. Implicitly, this assumes a consistent overall mean SED \citep{Richards_2006}, which could again be a source of bias between heterogeneously selected samples. However, as XQz5 is a well-defined sample with high completeness at the bright end, we make these comparisons in order to examine relative distributions of other quasar samples compared to our sample. We include our 0.75 anisotropy correction factor consistently in our comparisons to literature estimates of bolometric luminosity.

We first construct a comparison to the \citetalias{Trakhtenbrot_2011} sample in Figure \ref{fig:Mbh-Lbol}, using their \mgii\ measurements that we have rescaled to the \citet{Shen_2011} calibration. Both XQz5 and the flux-limited \citetalias{Trakhtenbrot_2011} sample constructed from SDSS DR6 are biased towards high Eddington ratios with similar Eddington ratio distributions. However, the XQz5 targets are measured to be both more luminous and have higher mass black holes than \citetalias{Trakhtenbrot_2011}. Both samples occupy the high mass and bright end of the subsample of SDSS DR16Q with well-constrained continuum luminosities and \mgii-based mass measurements between z = 0.35 -- 2.3. In total, we find 11 targets in XQz5 with mildly super-Eddington luminosity.

Using SDSS DR16Q quasars with \mgii\ measurements as the parent sample, we construct a luminosity-matched sample by collecting two quasars with the closest matching luminosity to each quasar in XQz5, resulting in a subsample composed of 164 quasars in the redshift range $0.6 < z < 2.0$. The median $\rm L_{\rm{bol}}$ difference of a selected SDSS DR16Q quasar and its associated XQz5 quasar is $\sim0.19$ dex. Despite this, due to the scarcity of high luminosity SDSS quasars, the exact luminosity distribution of XQz5 could not be reproduced. In the top panels of Figure \ref{fig:Comp_hists}, we compare the luminosity, black hole mass, and Eddington ratio distribution between XQz5, the luminosity-matched DR16Q subsample, and the WISSH sample \citep{Bischetti_2017_WISSH, Vietri_2018}, which is a high-luminosity SDSS and WISE-selected set of quasars across a wide range of redshifts ($1.8 < z < 4.7$) designed to reflect the most luminous known AGN in the universe. The black hole parameters of the WISSH sample are typically derived from the \civ\ emission line and $\rm L(1350)$, which are inter-calibrated to the results from the \mgii\ and \hbeta\ line but contribute additional scatter due to their weak correlation with other virial mass estimators \citep[e.g.,][]{baskin_laor_2005, Shen_2012}. Nevertheless, we find in Figure \ref{fig:Comp_hists}, that both of the luminosity-matched lower redshift samples host black holes with higher mass than in XQz5. Statistical analysis of the XQz5 and lower-redshift mass distributions with the Kolmogorov-Smirnov (KS) test yields p-values less than 1\%, indicating a statistically significant difference in the parent population. This also implies that ultraluminous quasars observed at lower redshift are generally less active with lower Eddington ratios than their counterparts at higher redshift, which is in concordance with recent literature \citep[e.g.,][]{Yang_2021, Farina_2022}. The mean Eddington ratio difference between XQz5 and the two lower-redshift samples is $\Delta\lambda_{\rm{Edd}} \sim0.3$

When comparing XQz5 with high-quality brightness-selected X-shooter quasar samples (XQ-100 and E-XQR-30+) in the bottom panels of Figure \ref{fig:Comp_hists}, we find all samples to be distributed towards high Eddington ratios, suggesting that all three samples are composed of rapidly accreting massive quasars. The mean(median) Eddington ratio of XQz5 is $\sim0.59(0.58)$, which is the highest mean(median) value of the three samples. The E-XQR-30+ sample has a long low-luminosity tail which consists entirely of literature quasars compiled in the \citet{Fan_2022} database. Unlike XQz5 and XQ-100, these quasars are not preferentially selected from the bright quasar population. By excluding the low luminosity tail of E-XQR-30+ such that the distribution better reflects the brightest known quasars at $5.8 < z < 6.6$, we find that the three Eddington ratio distributions are not statistically different as measured by the KS test.

In this comparison, we find XQz5 to be the brightest X-shooter quasar sample by proportion and by number of quasars brighter than $\log\left(L_{\rm{bol}}\rm{[erg s^{-1}]}\right) = 47.3$. As both XQz5 and XQR-30 are selected from the brightest known quasars at the time each sample was constructed, they are highly complete at the bright end of the quasar luminosity function. In particular, the completeness of the parent sample suggests that it is unlikely that a sample of brighter targets would exist within the redshift range of XQz5.

From XQR-30 at $z\sim6.1$ to the later cosmic epoch of XQz5 at $z\sim5.0$, we observe a significant increase in luminosity with decreasing redshift and a milder boost in black hole mass over a period of $\sim260$ Myr. This reflects an underlying population of quasars that are on a rising branch stage of their cosmic evolution, as their SMBHs have been feeding continuously from an earlier seed. A forthcoming paper will model and examine the exponential mass growth of quasars at this redshift range. 

\section{Summary and Conclusion} \label{sec:conclusion}

In this study, we described XQz5, a sample composed of 83 quasars with spectroscopic redshifts $4.5 < z < 5.3$. Selected from the brightest known quasars in a Southern sample of unprecedented completeness \citep{Onken_2022_QLF}, we have compiled high quality, moderate resolution spectra in the observed frame optical and near-infrared. The median reduced and post-processed SNR = 27.2 measured per 50 km s$^{-1}$ resolution element between 2700--2750\AA\ and 2850--2900\AA. Using the near-infrared observations, we examine the broad \mgii\ emission-line visible in rest-frame UV spectra and we measure properties of the black hole from the emission-line and the quasar continuum using established virial relations \citep{Shen_2011}. The main results are as follows:

\begin{itemize}
    \item By measuring the \mgii\ line present in all 83 quasar spectra and adopting virial relations from \citet{Shen_2011}, we find that the black hole mass range of our sample is $\log{(\rm{M_{BH}}/\rm{M_\odot})} = 8.5-10.4$. We also estimate the bolometric luminosity by applying a fixed bolometric correction of $k_{\rm{3000}} = 5.15$ \citep{Richards_2006}, with an anisotropy correction factor of 0.75 \citep{Runnoe_2012, Lai_23_AD}, finding the majority of targets in our sample to occupy the bolometric luminosity range $\log{(\rm{L_{bol}}/\rm{erg\,s^{-1}})} = 47.0-48.0$, making it the brightest quasar sample among other high-redshift X-shooter surveys, XQ-100 \citep{Lopez_2016_XQ100} and XQR-30 \citep{Dodorico_2023_XQR30}. The luminosity of 10 quasars in our sample appear mildly super-Eddington.
    \item XQz5 is a brighter sample with a 0.6 dex higher median luminosity and 0.4 dex more massive black holes compared to \citetalias{Trakhtenbrot_2011}, a large sample at a similar redshift range from the literature. However, the Eddington ratio distributions occupy a similar range, with both distributions skewed towards high Eddington ratios.
    \item Compared to a lower-redshift SDSS DR16Q \citep{Wu_2022} subset with \mgii\ measurements, XQz5 quasars occupy the high-mass and high-luminosity tail of the distribution, but when a luminosity-matched sample is constructed, the SDSS DR16Q black holes are more massive, indicating that ultraluminous quasars observed at higher redshift are generally more active than their lower-redshift counterparts, in agreement with recent literature \citep[e.g.,][]{Yang_2021, Farina_2022}.
\end{itemize}

Due to the high completeness of this sample at the bright end of the quasar luminosity function, this sample has notable legacy significance, as it is improbable that a brighter sample of a similar size could be constructed at this redshift range. Furthermore, as a sample that bridges the gap between XQ-100 \citep{Lopez_2016_XQ100} and XQR-30 \citep{Dodorico_2023_XQR30}, this sample has widespread community value. We have made the reduced spectra available in the supplementary material and in a publicly accessible repository.

\section*{Acknowledgements}
We thank Manuela Bischetti and Chiara Mazzucchelli for providing access to the WISSH sample and the $z > 5.6$ sample, respectively. We also thank the authors of \citet{Vestergaard_2001}, \citet{Tsuzuki_2006}, \citet{Bruhweiler_Verner_2008}, \citet{Mejia-Restrepo_2016}, \citet{Boroson_Greene_1992}, and \citet{Park_2022} for producing and sharing their \feii\ emission templates. We also thank Jack Hon and Jinyi Yang for access to their archival WiFeS data.

S.L. is grateful to the Research School of Astronomy \& Astrophysics at Australian National University for funding his Ph.D. studentship.

CAO was supported by the Australian Research Council (ARC) through Discovery Project DP190100252.

This paper is based on observations made with ESO Telescopes at the La Silla Paranal Observatory under programme IDs 084.A-0574(A), 084.A-0780(B), 087.A-0125(A), 094.A-0793(A), 098.A-0111(A), 0100.A-0243(A), 0104.A-0410(A), 108.22H9.001, 109.23D1.001, and 109.23D1.002.

The national facility capability for SkyMapper has been funded through ARC LIEF grant LE130100104 from the Australian Research Council, awarded to the University of Sydney, the Australian National University, Swinburne University of Technology, the University of Queensland, the University of Western Australia, the University of Melbourne, Curtin University of Technology, Monash University and the Australian Astronomical Observatory. SkyMapper is owned and operated by The Australian National University's Research School of Astronomy and Astrophysics. The survey data were processed and provided by the SkyMapper Team at ANU. The SkyMapper node of the All-Sky Virtual Observatory (ASVO) is hosted at the National Computational Infrastructure (NCI). Development and support of the SkyMapper node of the ASVO has been funded in part by Astronomy Australia Limited (AAL) and the Australian Government through the Commonwealth's Education Investment Fund (EIF) and National Collaborative Research Infrastructure Strategy (NCRIS), particularly the National eResearch Collaboration Tools and Resources (NeCTAR) and the Australian National Data Service Projects (ANDS).

The Pan-STARRS1 Surveys (PS1) and the PS1 public science archive have been made possible through contributions by the Institute for Astronomy, the University of Hawaii, the Pan-STARRS Project Office, the Max-Planck Society and its participating institutes, the Max Planck Institute for Astronomy, Heidelberg and the Max Planck Institute for Extraterrestrial Physics, Garching, The Johns Hopkins University, Durham University, the University of Edinburgh, the Queen's University Belfast, the Harvard-Smithsonian Center for Astrophysics, the Las Cumbres Observatory Global Telescope Network Incorporated, the National Central University of Taiwan, the Space Telescope Science Institute, the National Aeronautics and Space Administration under Grant No. NNX08AR22G issued through the Planetary Science Division of the NASA Science Mission Directorate, the National Science Foundation Grant No. AST-1238877, the University of Maryland, Eotvos Lorand University (ELTE), the Los Alamos National Laboratory, and the Gordon and Betty Moore Foundation.

The VISTA Hemisphere Survey data products served at Astro Data Lab are based on observations collected at the European Organisation for Astronomical Research in the Southern Hemisphere under ESO programme 179.A-2010, and/or data products created thereof.

This work is based in part on data obtained as part of the UKIRT Infrared Deep Sky Survey. This work is also based in part on observations obtained at the Southern Astrophysical Research (SOAR) telescope, which is a joint project of the Minist\'{e}rio da Ci\^{e}ncia, Tecnologia e Inova\c{c}\~{o}es (MCTI/LNA) do Brasil, the US National Science Foundation’s NOIRLab, the University of North Carolina at Chapel Hill (UNC), and Michigan State University (MSU).

This publication has made use of data from the VIKING survey from VISTA at the ESO Paranal Observatory, programme ID 179.A-2004. Data processing has been contributed by the VISTA Data Flow System at CASU, Cambridge and WFAU, Edinburgh.

This publication makes use of data products from the Two Micron All Sky Survey, which is a joint project of the University of Massachusetts and the Infrared Processing and Analysis Center/California Institute of Technology, funded by the National Aeronautics and Space Administration and the National Science Foundation.

This publication makes use of data products from the Wide-field Infrared Survey Explorer, which is a joint project of the University of California, Los Angeles, and the Jet Propulsion Laboratory/California Institute of Technology, and NEOWISE, which is a project of the Jet Propulsion Laboratory/California Institute of Technology. WISE and NEOWISE are funded by the National Aeronautics and Space Administration.

Software packages used in this study include \texttt{Numpy} \citep{Numpy_2011}, \texttt{Scipy} \citep{Scipy_2020}, \texttt{Astropy} \citep{Astropy_2013}, \texttt{PypeIt} \citep{Prochaska_2020_Pypeit}, \texttt{Specutils} \citep{specutils_2022}, \texttt{SpectRes} \citep{Carnall_2017}, and \texttt{Matplotlib} \citep{Matplotlib_2007}.

\section*{Data Availability}
The data underlying this article will be shared on reasonable request to the corresponding author. Reduced spectra can be downloaded from this GitHub repository: \href{https://github.com/samlaihei/XQz5}{https://github.com/samlaihei/XQz5}.



\bibliographystyle{mnras}
\bibliography{bibliography} 




\appendix
\section{Additional Figures} \label{appendix:additional_tables_figs}
The following Figure \ref{fig:FigGrid} presents fits of the pseudo-continuum and \mgii\ line for all 83 quasars in XQz5. \mgii\ line models derived using the \citet{Tsuzuki_2006} \feii\ template are shown, and the remaining diagnostic figures can be located in the GitHub repository. The reported fit parameters include the systematic errors from varying the template. Figure \ref{fig:sample_spectrum} shows the spectrum of SDSS J001115.23$+$144601.8 at redshift $z=4.96$ from observed frame $\sim$6000\AA--22000\AA. The full wavelength coverage of our data extends beyond the plotted region. The underlying spectrum shown in grey is the reduced data which is made available online in the GitHub repository. The coloured spectrum shows the resulting spectrum after the post-processing routine described in Section \ref{sec:post-processing}, which is applied to rest-frame 1050--3600\AA. Several prominent broad lines are labeled, and the error spectrum is presented in the bottom of the panel in dark grey. Similar figures are created for all quasars in this sample and are made available in the online repository.

\renewcommand{\thefigure}{A1}
\begin{figure*}
	\includegraphics[width=0.9\textwidth]{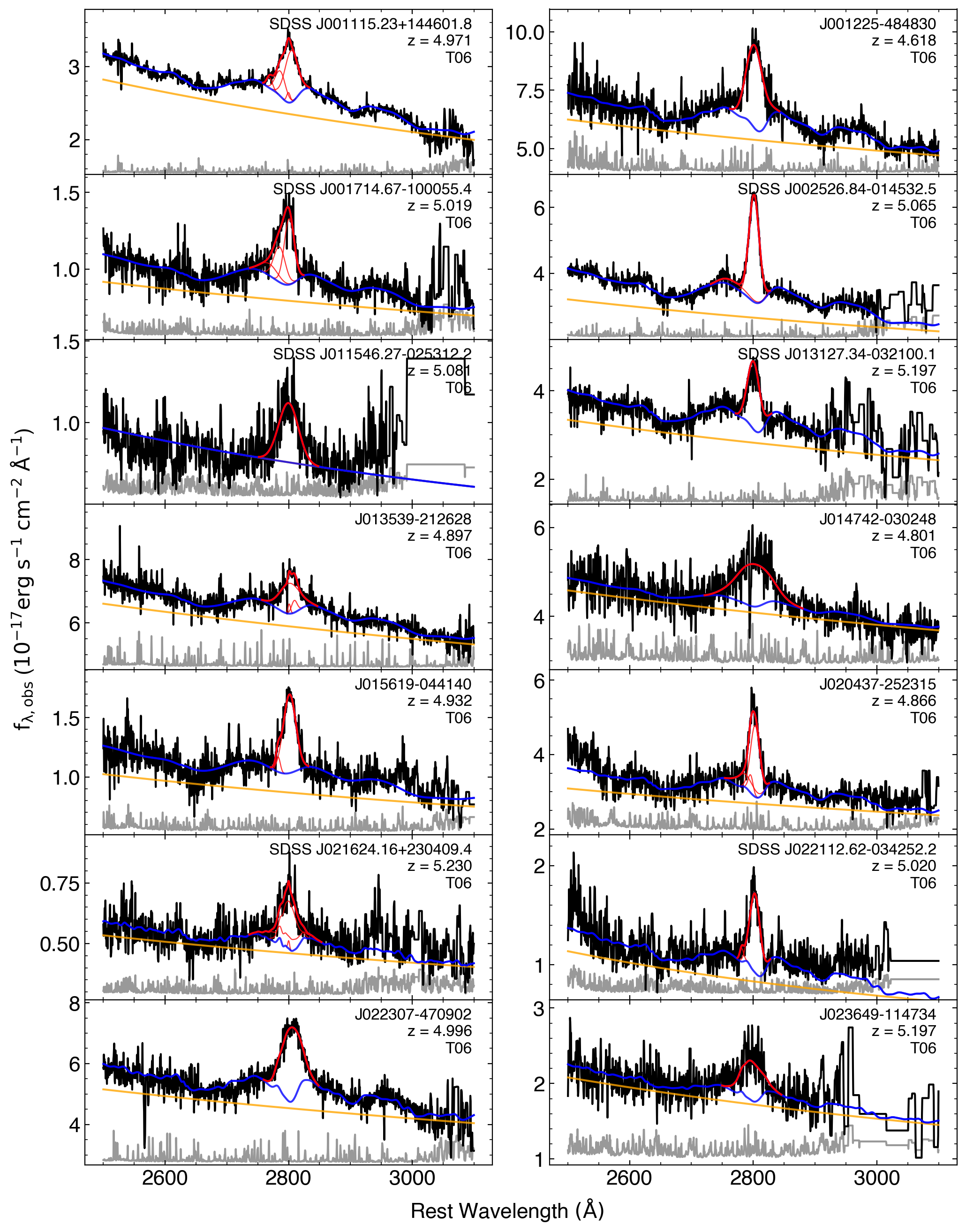}
    \caption{Rest-frame spectra of all ultraluminous quasar sources presented in this study, ordered by increasing right ascension. Within each panel, the data is plotted in black with the best-fit power-law continuum in orange, pseudo-continuum in blue, and \mgii\ line model in red (with the individual Gaussian components shown as narrower red lines). The error spectrum, presented in grey, is shifted vertically such that the bottom of the panel represents zero flux error. We show an example line model for the \citet{Tsuzuki_2006} template with the rest of the line models available as online supplementary material.}
    \label{fig:FigGrid}
\end{figure*}

\begin{figure*}
	\includegraphics[width=0.9\textwidth]{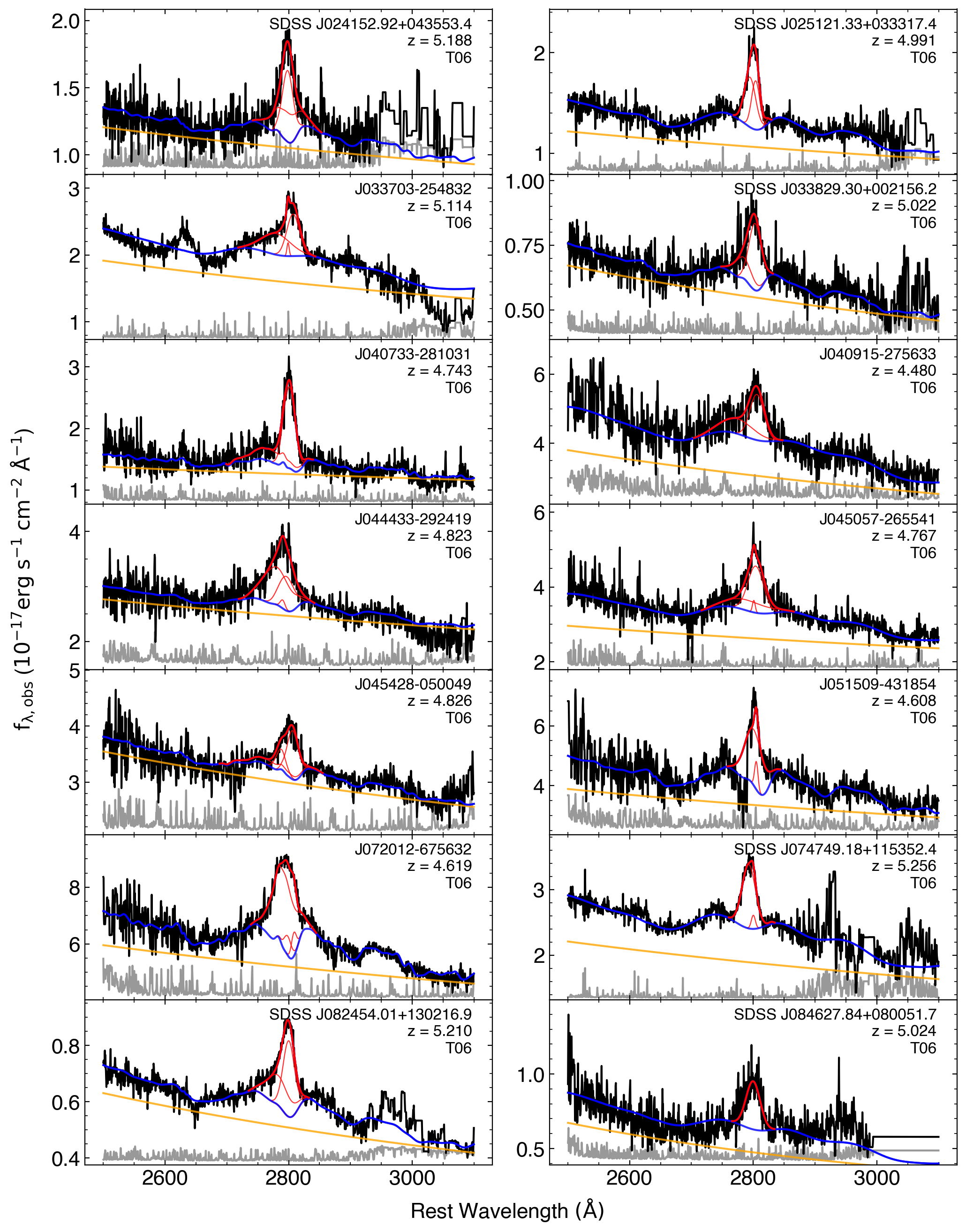}
    \caption{(Continued)}
\end{figure*}

\begin{figure*}
	\includegraphics[width=0.9\textwidth]{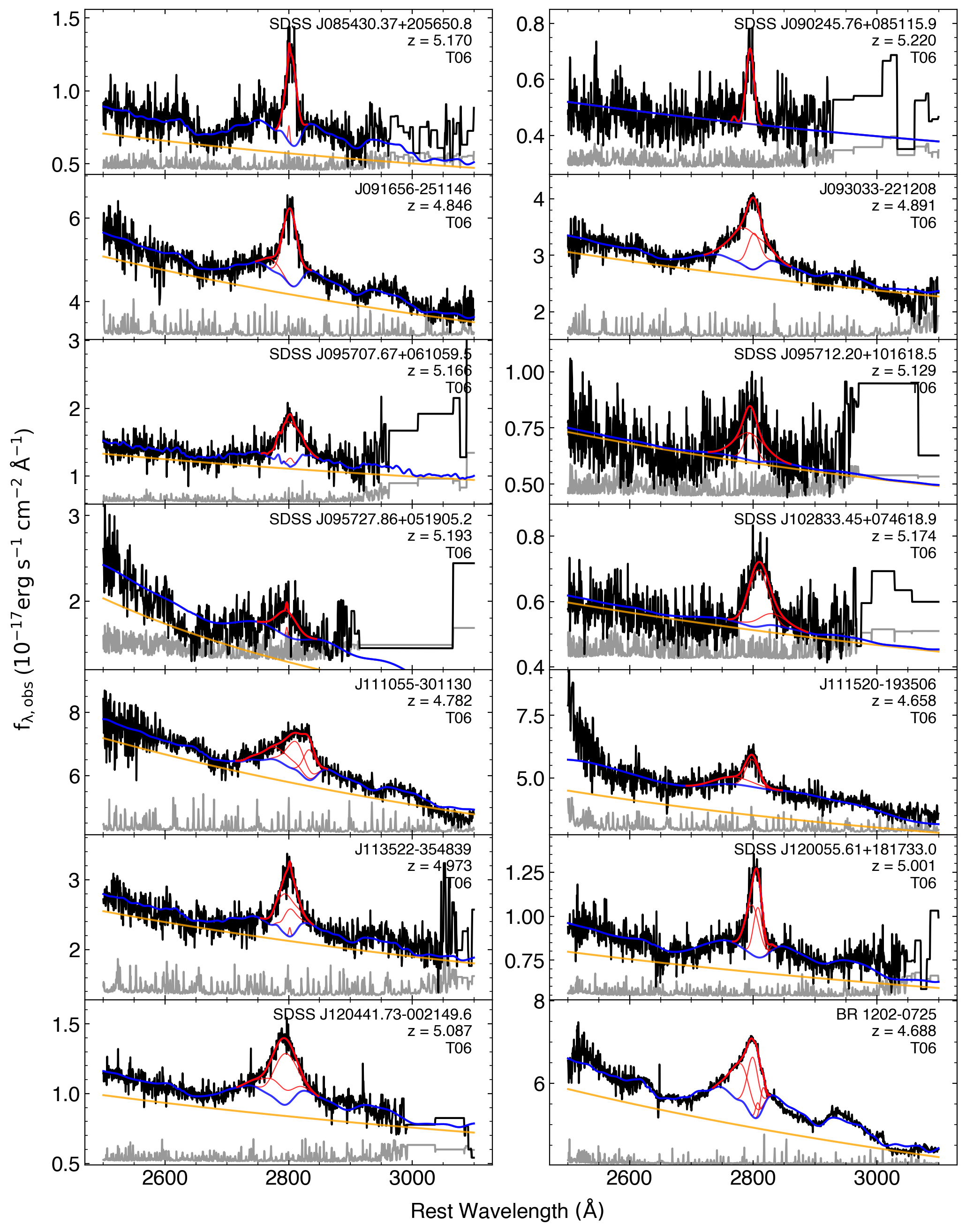}
    \caption{(Continued)}
\end{figure*}

\begin{figure*}
	\includegraphics[width=0.9\textwidth]{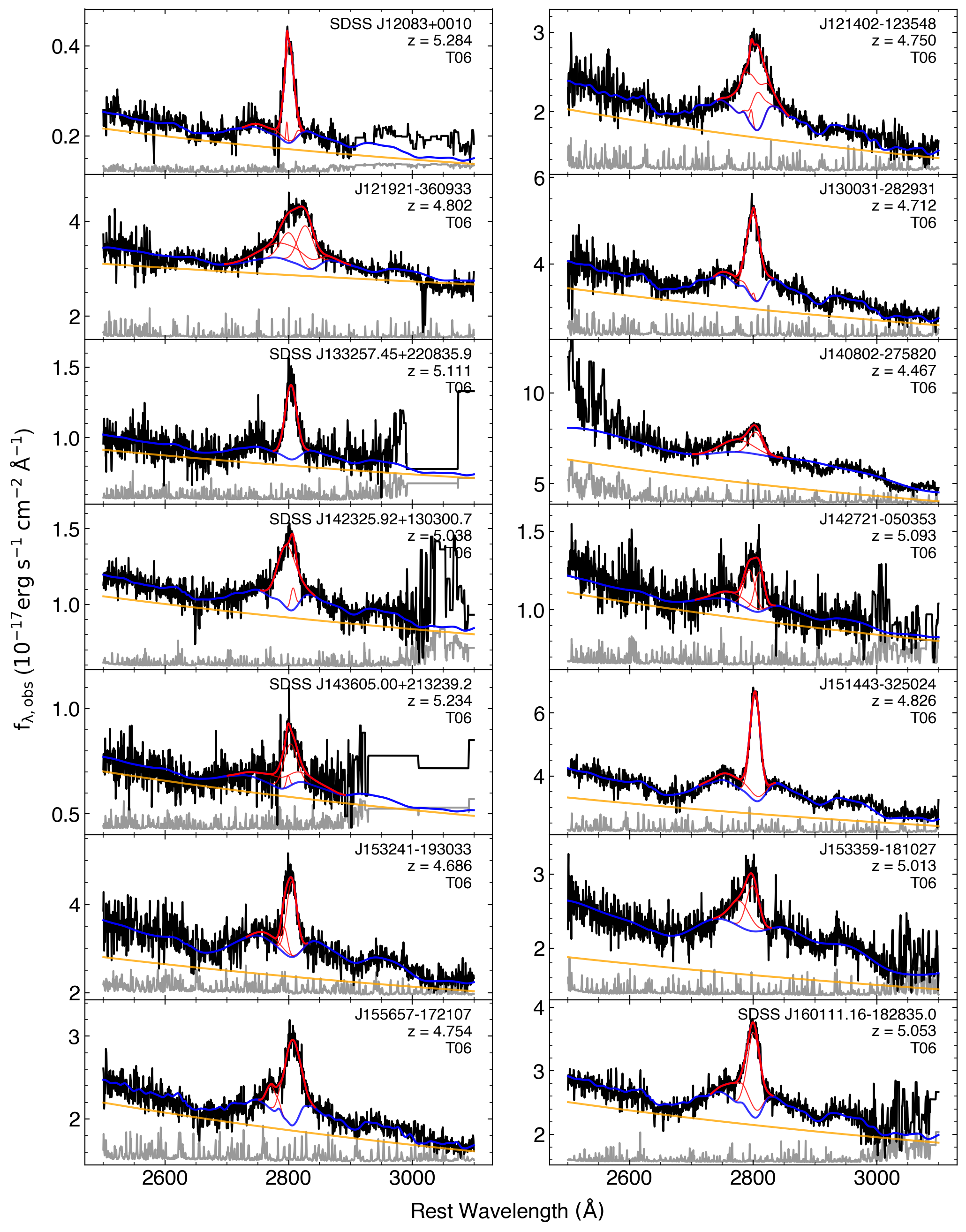}
    \caption{(Continued)}
\end{figure*}

\begin{figure*}
	\includegraphics[width=0.9\textwidth]{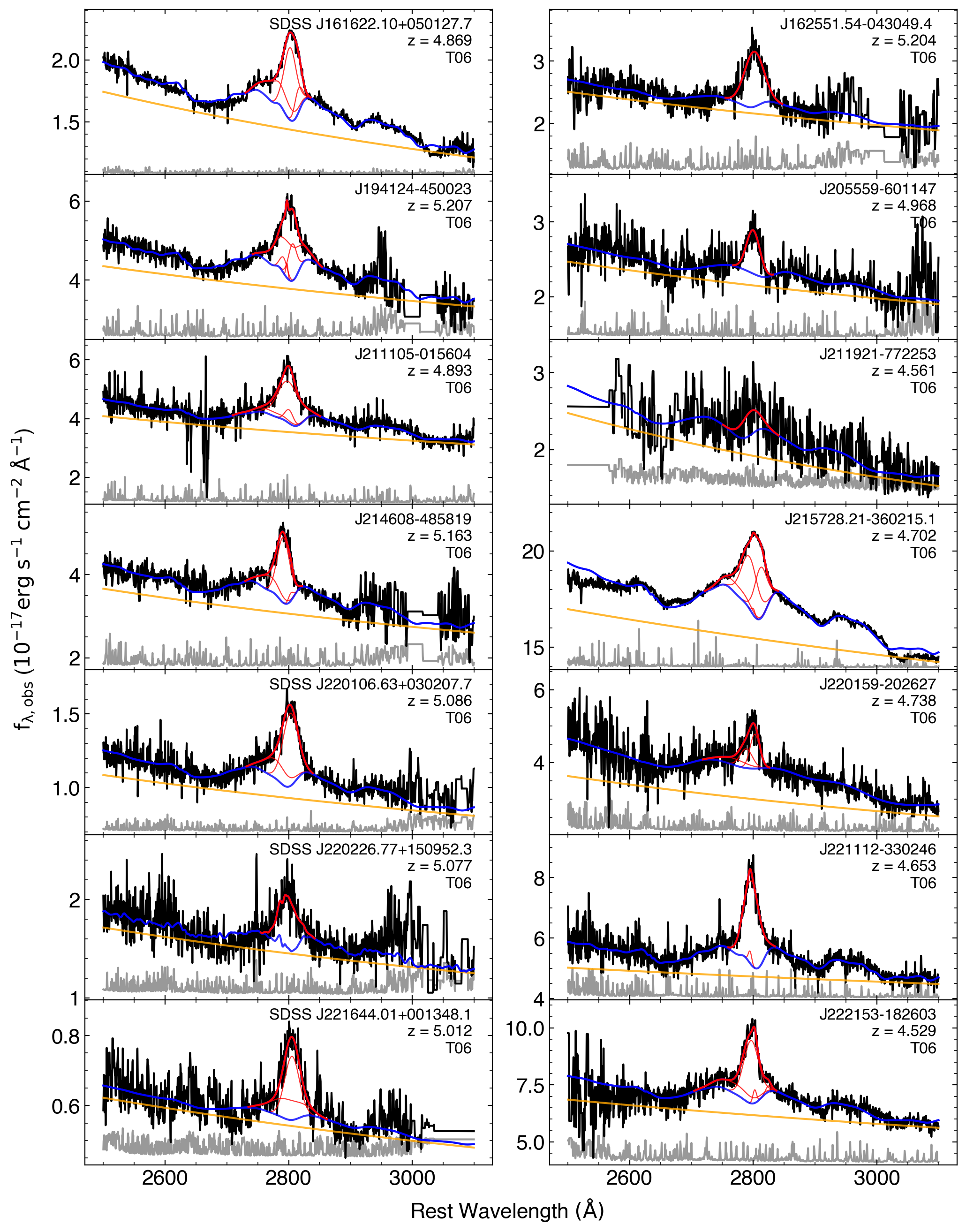}
    \caption{(Continued)}
\end{figure*}

\begin{figure*}
	\includegraphics[width=0.9\textwidth]{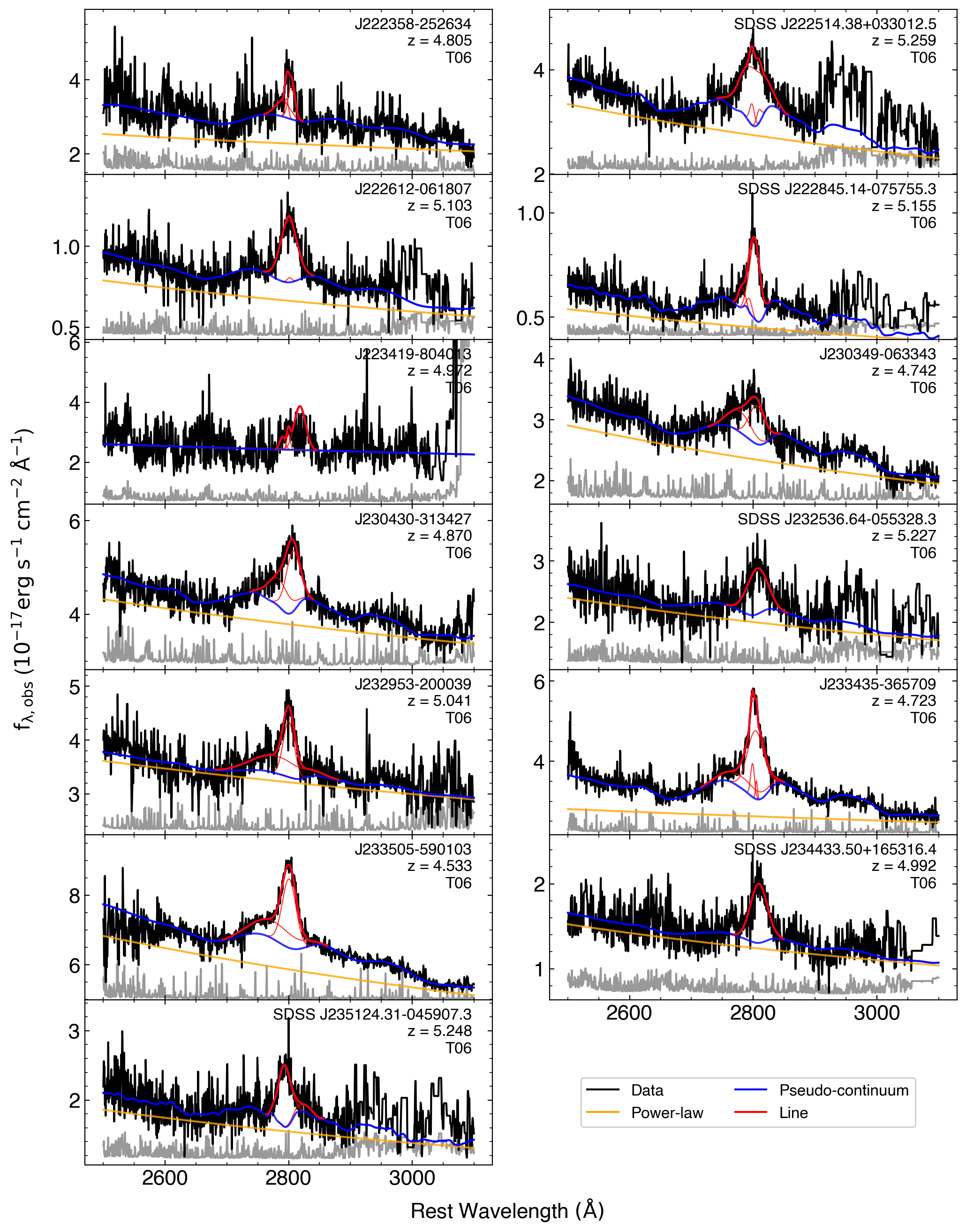}
    \caption{(Continued)}
\end{figure*}

\renewcommand{\thefigure}{A\arabic{figure}}
\setcounter{figure}{1}

\begin{figure*}
	\includegraphics[width=0.9\textwidth]{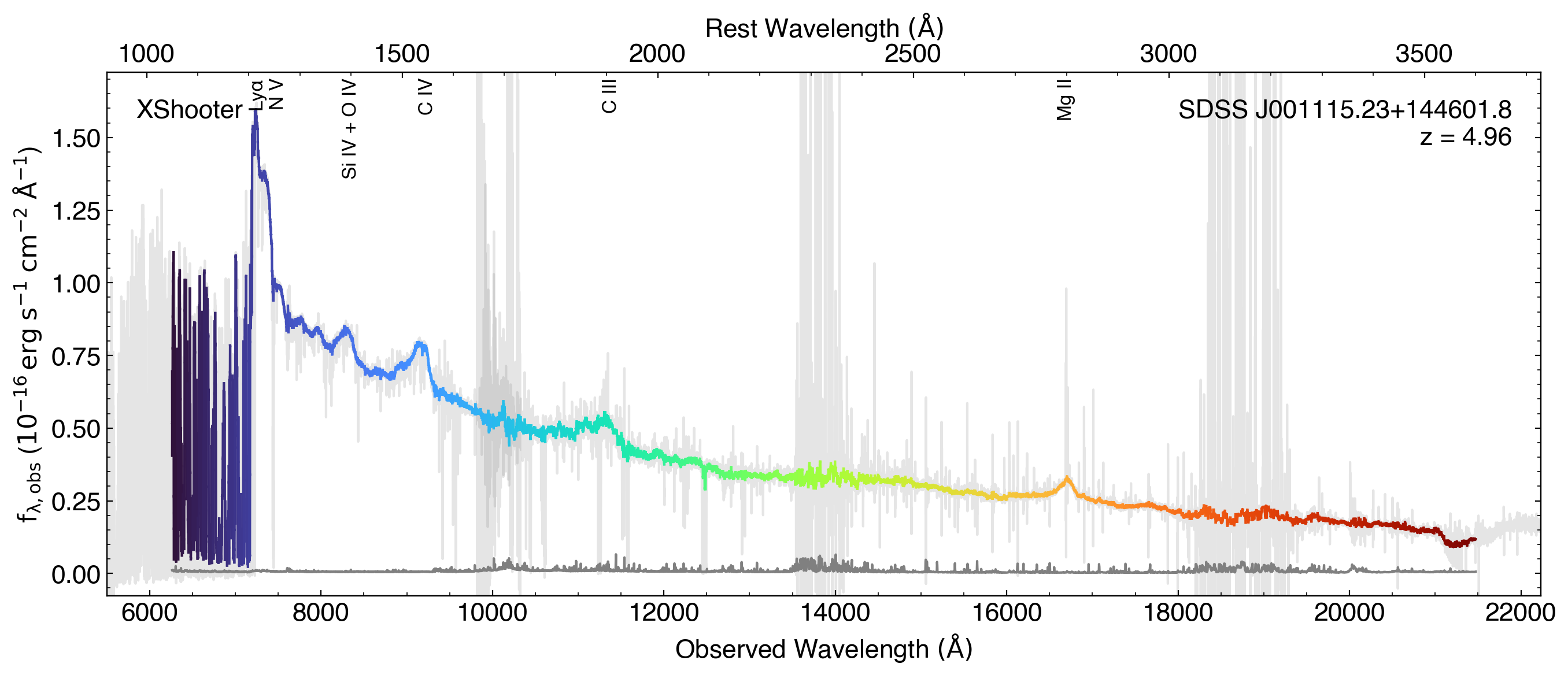}
    \caption{Spectrum of SDSS J001115.23$+$144601.8 at redshift $z=4.96$ from observed frame $\sim$6000\AA--22000\AA. The grey data underlying the coloured spectrum show the reduced data that is made available as online supplementary material, while the coloured spectrum shows the resulting data after applying the post-processing steps of Section \ref{sec:post-processing}. The error spectrum is presented in dark grey at the bottom of the panel.}
    \label{fig:sample_spectrum}
\end{figure*}


\bsp	
\label{lastpage}
\end{document}